\documentclass[aps,twocolumn,pre]{revtex4}

\usepackage{pifont}

\usepackage{amsfonts}

\usepackage{pifont}
\newcommand{\tickYess}{\ding{52}}
\newcommand{\tickNo}{\hspace{1pt}\ding{55}}

\usepackage{latexsym}
\usepackage[pdftex]{graphicx}
\begin{document}

\title{Social features of online networks: the strength of intermediary ties in online social media}

\author{Przemyslaw A. Grabowicz$^1$}\email{pms@ifisc.uib-csic.es}
\author{ Jos\'e J. Ramasco$^1$}\email{jramasco@ifisc.uib-csic.es} \author{Esteban Moro$^{2,3}$} 
\author{Josep M. Pujol$^{4,5}$} 
\author{Victor M. Eguiluz$^1$}

\affiliation{$^1$Instituto de F\'{\i}sica Interdisciplinar y Sistemas Complejos IFISC (CSIC-UIB), Palma de Mallorca, Spain\\
$^2$Instituto de Ingenier\'{\i}a del Conocimiento, Universidad Aut\'onoma de Madrid, Madrid, Spain\\
$^3$Instituto de Ciencias Matem\'aticas CSIC-UAM-UC3M-UCM, Departamento de Matem\'aticas \& GISC, Universidad Carlos III de Madrid, Legan\'es, Spain \\
$^4$Telef\'onica Research, 08019 Barcelona, Spain\\
$^5$3scale Networks, 08018 Barcelona, Spain
}

\begin{abstract}
An increasing fraction of today social interactions occur using online social media as communication channels. Recent worldwide events, such as social movements in Spain or revolts in the Middle East, highlight their capacity to boost people coordination. Online networks display in general a rich internal structure where users can choose among different types and intensity of interactions. Despite of this, there are still open questions regarding the social value of online interactions. For example, the existence of users with millions of online friends sheds doubts on the relevance of these relations. In this work, we focus on Twitter, one of the most popular online social networks, and find that the network formed by the basic type of connections is organized in groups. The activity of the users conforms to the landscape determined by such groups. Furthermore, Twitter's distinction between different types of interactions allows us to establish a parallelism between online and offline social networks: personal interactions are more likely to occur on internal links to the groups (the weakness of strong ties), events transmitting new information go preferentially through links connecting different groups (the strength of weak ties) or even more through links connecting to users belonging to several groups that act as brokers (the strength of intermediary ties).
\end{abstract}
\maketitle

\section{Introduction}

There exists an open discussion on the validity of online interactions as indicators of real social activity~\cite{cummings02,dijk06,watts07,avnit09,alex09,lazer09}. Most of the online social networks incorporate several types of user-user interactions that satisfy the need for different level of involvement or relation intensity between users~\cite{lewis08,honeycutt09,szell10,gruzd11,ferrara11}. The cost of establishing the cheapest relation is usually very low, and it requires the acceptation or simply the notification to the targeted user. These connections can accumulate due to the asymmetric social cost of cutting and creating them, and pile up to the astronomic numbers that capture popular imagination~\cite{avnit09}. If the number of connections increases to the thousands or the millions, the amount of effort that a user can invest into the relation that each link represents must fall to near zero. Does this mean that online networks are irrelevant for understanding social relations, or for predicting where higher quality activity (e.g., personal communications, information transmission events) is taking place? By analyzing the clusters of the network formed by the cheapest connections between users of Twitter, we show that even this network bears valuable information on the localization of more personal interactions between users. Furthermore, we are able to identify some users that act as brokers of information between groups.

The theory known as {\it the strength of weak ties} proposed by Granovetter~\cite{granovetter73} deals with the relation between structure, intensity of social ties and diffusion of information in offline social networks. It has raised some interest in the last decades~\cite{granovetter73,csermely06,onnela07,esteban11} and its predictions have been checked in a mobile phone calls dataset~\cite{onnela07}. On one hand, a tie can be characterized by its strength, which is related to the time spend together, intimacy and emotional intensity of a relation. Strong ties refer to relations with close friends or relatives, while weak ties represent links with distant acquaintances. On the other hand, a tie can be characterized by its position in the network. Social networks are usually composed of groups of close connected individuals, called communities, connected among them by long range ties known as bridges. A tie can thus be internal to a group or a bridge. Grannoveter's theory predicts that weak ties act as bridges between groups and are important for the diffusion of new information across the network, while strong ties are usually located at the interior of the groups. Burt's work~\cite{burt05} later emphasizes the advantage of connecting different groups (bridging structural holes) to access novel information due to the diversity in the sources. More recent works, however, point out that information propagation may be dependent on the type of content transmitted~\cite{centola07,centola10} and on a \textit{diversity-bandwidth tradeoff}~\cite{aral11}. The bandwidth of a tie is defined as the rate of information transmission per unit of time. Aral et al.~\cite{aral11} note that weak ties interact infrequently, therefore have low bandwidth, whereas strong ties interact more often and have high bandwidth. The authors claim that both diversity and bandwidth are relevant for the diffusion of novel information. Since both are anticorrelated, there has to be a tradeoff to reach an optimal point in the propagation of new information. They also suggest that strong ties may be important to propagate information depending on the structural diversity, the number of topics and the dynamic of the information. Due to the different nature of online and offline interactions, it is not clear whether online networks organize following the previous principles. Our aim in this work is to test if these theories apply also to online social networks.

Online networks are promising for such studies because of the wide data availability and the fact that different type of interactions are explicitly separated: e.g.,  information  diffusion events are distinguished from more personal communications. Diffusion events are implemented as a system option in the form of \textit{share} or \textit{repost} buttons with which it is enough to single-click on a piece of information to rebroadcast it to all the users' contacts. This is in contrast to personal communications and information creation for which more effort has to be invested to write a short message and (for personal communication) to select the recipient. All these features are present in Twitter, which is a micro-blogging social site. The users, identified with a username, can write short messages of up to $140$ characters (tweets) that are then broadcasted to their followers. When a new follower relation is established, the targeted user is notified although his or her explicit permission is not required. This is the basic type of relation in the system~\cite{java07,krishna08,huberman08}, which generates a directed graph connecting the users: the follower network. After some time of functioning, some peculiar behaviors started to extend among Twitter users leading to the emergence of particular types of interactions. These different types of interactions have been later implemented as part of Twitter's system~\cite{blog2}. \textit{Mentions} (tweets containing @username) are messages which are either directed only to the corresponding user or mentioning the targeted user as relevant to the information expressed to a broader audience. A \textit{retweet} (RT @username) corresponds to content forward with the specified user as the nominal source. In contrast to the normal tweets, mentions usually include personal conversations or references~\cite{honeycutt09} while retweets are highly relevant for the viral propagation of information~\cite{galuba10}. This particular distinction between different types of interactions qualifies Twitter as a perfect system to analyze the relation between topology, strength of social relation and information diffusion in online social networks.

The properties of the follower network have been extensively analyzed especially in relation to its topological structure, propagation of information, homophily, tie formation and decay, etc~\cite{kwak10,mendoza10,jakob10,asur11,romero10,pujol10,borge11}. Finding users with thousands or even millions of followers is not exceptional~\cite{avnit09}, so the question is whether the structure of the follower network carries any information on where personal relations (mentions) or information transmission events (retweets) take place. To answer this question, we first analyze a sample of the follower network with clustering-detection algorithms and identify a set of groups. Our dataset is a sample of the network containing $2\,408\,534$ users connected with $48\,776\,888$ follower relations, as well as the tweets, retweets, mentions, and was gathered through the Twitter API  during November and December of $2008$~\cite{pujol10, pujol2, erramili} (see the Methods Section for further detail). Whether the clusters we identify are traces of underlying social groups (online or offline) is a question we cannot answer with the available information. We follow an alternative path by checking the correlation between the location of the personal conversations (mentions) and information diffusion events (retweets) and the structural properties of the link bearing those activities with respect to the detected groups in the network. Note that we consider mentions and retweets to happen always on follower links. This allow us to describe user activity in terms of the detected groups. 

\begin{figure}[b]
\begin{center}
\includegraphics[width=8.6cm]{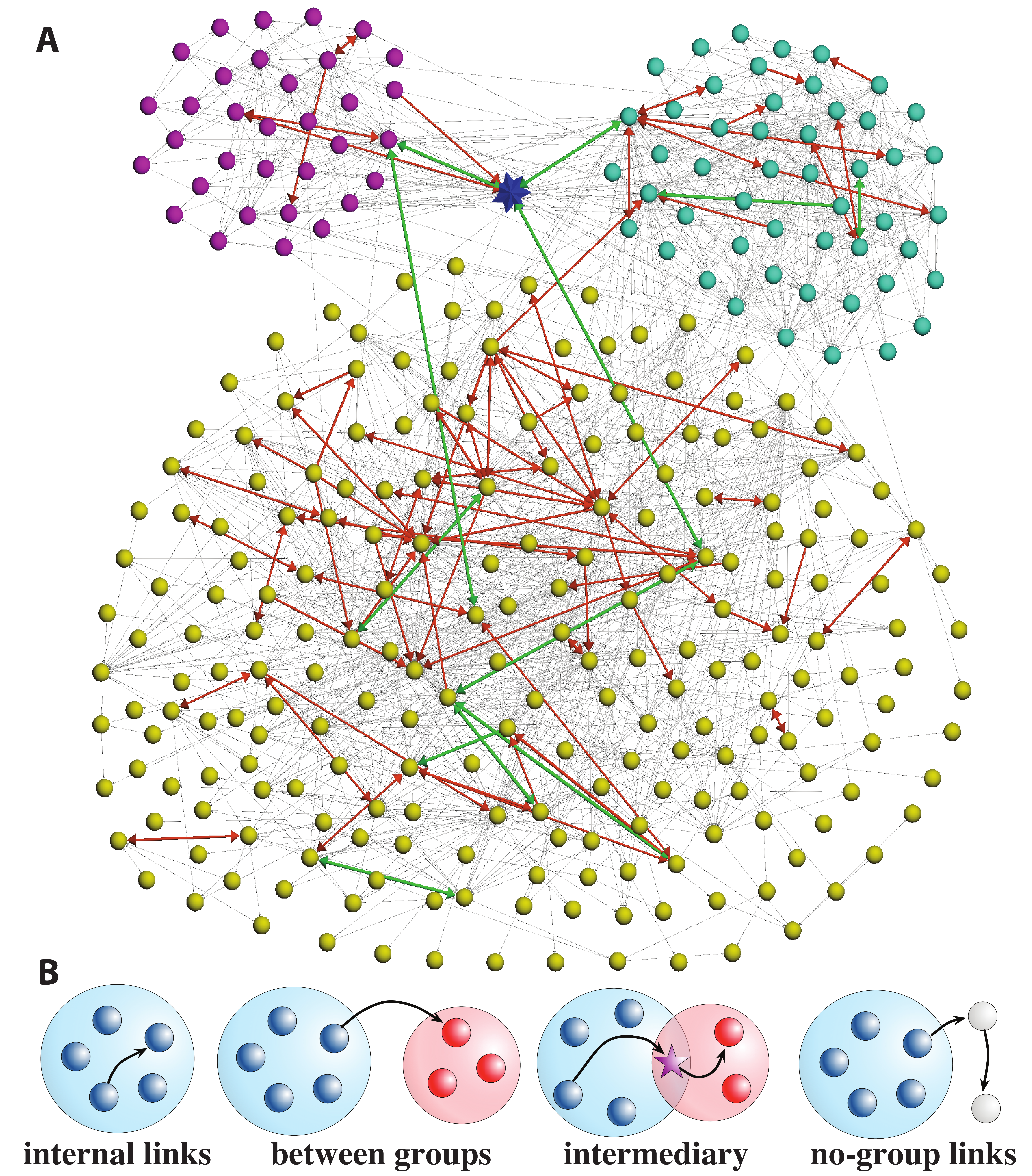}
\caption{Groups and links. (A) Sample of Twitter network: nodes represent users and links, interactions. The follower connections are plotted as gray arrows, mentions in red, and retweets in green. The width of the arrows is proportional to the number of times that the link has been used for mentions. We display three groups (yellow, purple and turquoise) and a user (blue star) belonging to two groups. (B) Different types of links depending on their position with respect to the groups' structure: internal, between groups, intermediary links and no-group links.}\end{center}
\end{figure}

\section{Results}

\subsection{Description of the groups}

Our first step is to identify the groups in the follower network. Clustering in large graphs is still a topic of very active research and many algorithms are available~\cite{santo10}. Due to the size, density, and directness of the follower network and in order to capture the possible inclusion of users in multiple groups or in none, we have used Oslom~\cite{oslom11,significance10} (see Methods). The analysis has also been performed with other clustering techniques~\cite{rosvall08,mcdaid10,raghavan,louvain,radatools}, reaching similar conclusions (see Setion $3$ in Supplementary Information [Figs. S6-S14 and Table S1]  for a detailed account on these results). We have detected $92,062$ groups, three of which are graphically depicted in Figure~1A with each sphere corresponding to a single user. In general, the links can be classified according to their position with respect to the user groups: internal, between groups, intermediary and links involving nodes not assigned to any group as shown in Figure~1B.

The statistics characterizing the groups and links are displayed in Figure~2. The group size distribution decays slowly for three orders of magnitude and does not show a characteristic group size (Figure~2A). For instance, the largest group contains around $10,000$ users. Also the number of groups each user belongs to shows high heterogeneity: $37.4 \%$ of the users has not been allocated to any group, while there exists a user belonging to more than $100$ groups (see Figure~2B). The percentage of links falling in the different types regarding the groups is depicted in Figure~2C. Although the non-classified users are $37\%$ of the total, the links connected to them are less than $6\%$ and the percentage is even lower for those with mentions or retweets. The most common type of connections is the between-group links. 
One may wonder if the algorithm for clusters detection is doing a good job when there is such a large proportion of between-group links. The clustering method is trying to find groups of mutually interconnected nodes that would be extremely rare in a randomized instance of the network, rather than optimizing the ratio between number of between-group and internal links. In Sections $1$ and $2$ of the Supplementary Information (Figs. S1-S5), this argument is further developed and the capacity of Oslom to detect planted communities is proved in a benchmark even in situations with a high ratio between the number of between-groups and internal links. Another relevant point to highlight is the different potential of each type of links to carry mentions and retweets. As it can be seen in the Figure~2C, the red bars for mentions in internal links and intermediary links almost double the abundance of links in the follower network in these categories. The links between groups, on the other hand, attract far less mentions.

\begin{figure}
\begin{center}
\includegraphics[width=8.6cm]{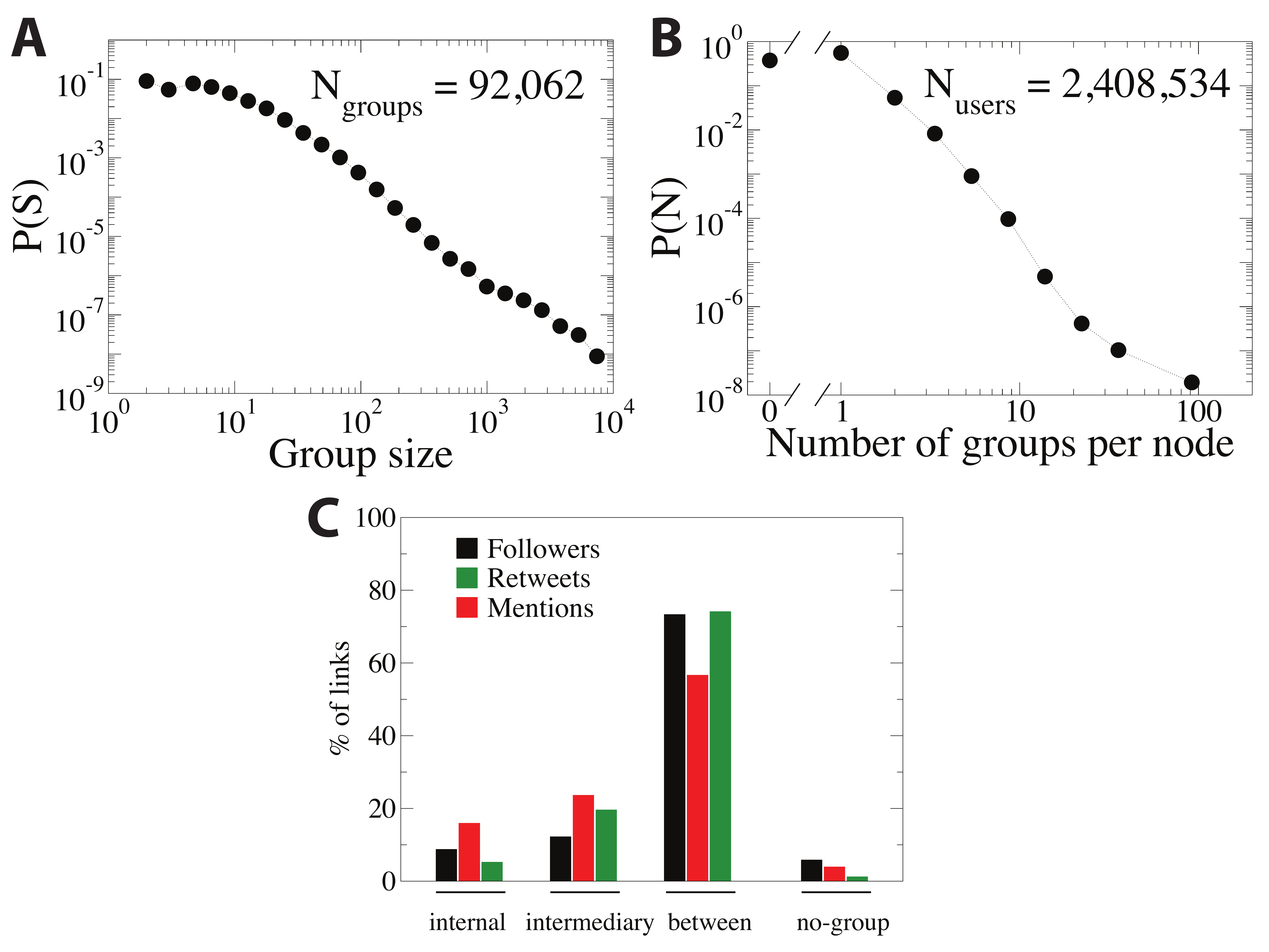}
\caption{ Group and link statistics. (A) Size distribution of the group. (B) Distribution of the number of groups to which each user is assigned. (C) Percentage of links of different types, e.g. follower links (black bars), links with mentions (red bars) or retweets (green bars), staying in particular topological localizations in respect to detected groups.}\end{center}
\end{figure}

\begin{figure}[t]
\begin{center}
\includegraphics[width=8.6cm]{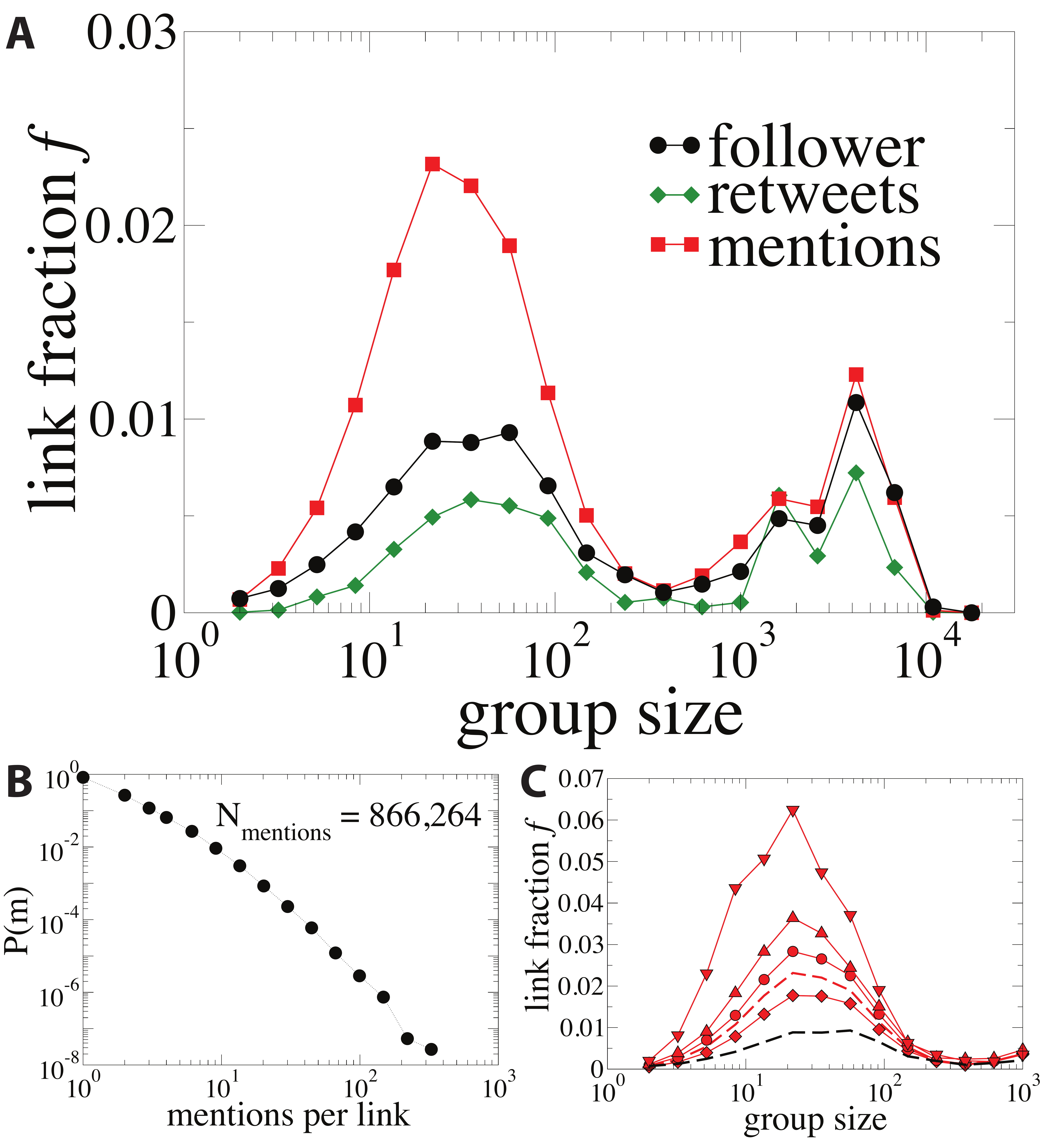}
\caption{Internal activity. (A) Fraction $f$ of internal links as a function of the group size in number of users. The curve for the follower network acts as baseline for mentions and retweets. Note that if mentions/retweets were randomly appearing over follower links then the red/green curve should match the black curve. (B) Distribution of the number of mentions per link. (C) Fraction of links with mentions as a function of their intensity. The dashed curves are the total for the follower network (black) and for the links with mentions (red). While the other curves correspond (from bottom to top) to fractions of links with: 1 non-reciprocated mention (diamonds), 3 mentions (circles), 6 mentions (triangle up) and more than 6 reciprocated mentions (triangle down).}\end{center}
\end{figure}

\subsection{The strength of ties}

Besides their location with respect to the groups, the links can be also characterized by their intensity. In Twitter mentions are typically used for personal communication, which establishes a parallelism between links with mentions and strength of social ties. The more mentions has been exchanged between two users, even more so if reciprocated, the stronger we consider the tie between them. We define intensity of a link as the number of mentions interchanged on it. Different predictors have been considered to estimate social tie strength~\cite{marsden84} including, for instance, time spent together~\cite{marsden84} or the duration of phone calls~\cite{onnela07}. We consider the intensity as an approximation to social strength given that writing a mention involves some effort and addresses only single targeted users. 

\subsection{Internal links} 

According to Granovetter's theory, one could expect the internal connections inside a group to bear closer relations. Mechanisms such as homophily~\cite{mcpherson01}, cognitive balance~\cite{heider58,newcomb61} or triadic closure \cite{granovetter73} 
favor this kind of structural configurations. Unfortunately, we have no means to measure the closeness of a user-user relation in a sociological sense in our Twitter dataset. However we can verify whether the link has been used for mentions, whether the interchange has been reciprocated or whether it has happened more than once.  We define the fraction $f_p^i$ of links with interaction $i$ in position $p$ with respect to the groups of size $s$ as
\begin{equation}
f_p^i(s)  = \frac{n_p^i(s)}{N^i} ,
\end{equation}
where $n_p^i(s)$ is the number of links with that type of interaction in position $p$ with respect to the groups of size $s$ and $N^i$ in the total number of links with interaction $i$.  The fractions $f_{internal}^i(s)$  reveals an interesting pattern as function of the group size as can be seen in Figure~3A. Note that the fraction of links in the follower network (black curve) is taken as the reference for comparison. Links with mentions are more abundant as internal links than the baseline follower relations for groups of size up to $150$ users. This particular value brings reminiscences of the quantity known as the Dunbar number~\cite{dunbar}, the cognitive limit to the number of people with whom each person can have a close relationship and that has recently been discussed in the context of Twitter~\cite{bruno11}. Although we have identified larger groups, the density of mentions is similar to the density of links in the follower network.  In addition, the distribution of the number of times that a link is used (intensity) for mentions is wide, which allows for a systematic study of the dependence of intensity and position (see Figure~3B). The more intense (or reciprocated) a link with mentions is, the more likely it becomes to find this link as internal (Figure~3C). This corresponds to Granovetter expectation that the stronger the tie is the higher number of mutual contacts of both parties it

\begin{figure}[t]
\begin{center}
\includegraphics[width=8.6cm]{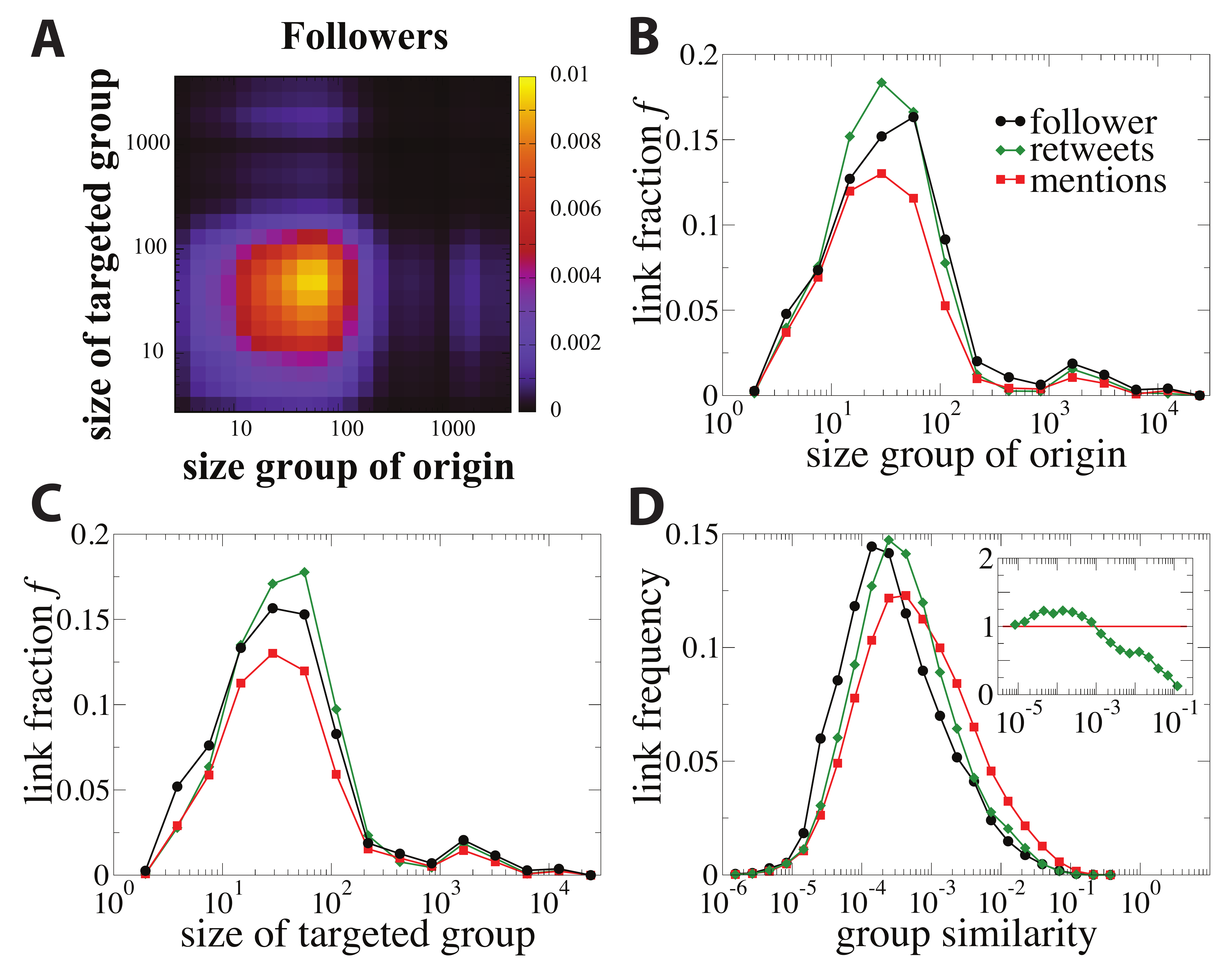}
\caption{ Group-group activity. (A) Distribution of the number of links in the follower network between groups as a function of the size of the groups. (B) Fractions $f$ of links of the different types (follower, with mentions and with retweets) as a function of the size of the group at the link origin, and (C) at the targeted group. (D) Frequency of between-group links as a function of the group-group similarity for the different type of links. In the inset, ratio between the frequency of links with retweets and with mentions.}
\end{center}
\end{figure}

\subsection{Links between groups} 

The next question to consider is the characteristics of links between groups. These links occur mainly between groups containing less than $200$ users (Figure~4A-C). However, their frequency depends on the quality of the links (if they bear mentions or retweets). While links with mentions are less abundant than the baseline, those with retweets are slightly more abundant. According to the strength of weak ties theory~\cite{granovetter73,onnela07,esteban11,burt05}, weak links are typically connections between persons not sharing neighbors, being important to keep the network connected and for information diffusion. We investigate whether the links between groups play a similar role in the online network as information transmitters. The actions more related to information diffusion are retweets~\cite{galuba10} that show a slight preference for occurring on between-group links (Figures~4B and ~4C). This preference is enhanced when the similarity between connected groups is taken into account. We define the similarity between two groups, A and B, in terms of the Jaccard index of their connections:
\begin{equation}
similarity(A,B) = \frac{|\cap \mbox{links of $A$ and $B$}| }{|\cup \mbox{links of $A$ and $B$}|}.
\end{equation}
The similarity is the overlap between the groups' connections and it estimates network proximity of the groups. The general pattern is that links with mentions more likely occur between close groups and retweets occur between groups with medium similarity (Figure~4D). Mentions as personal messages are typically exchanged between users with similar environments, what is predicted by the strength of weak ties theory. Links with retweets are related to information transfer and the similarity of the groups between which they take place should be small according to the Granovetter's theory. The results show that the most likely to attract retweets are the links connecting groups that are neither too close nor too far. This can be explained with Aral's theory about the trade-off between diversity and bandwidth: if the two groups are too close there is no enough diversity in the information, while if the groups are too far the communication is poor. These trends are not dependant on the size of the considered groups (see Section 3 [Figs S6-S14 and Table S1] in the Supplementary Information).

\subsection{Intermediary links} 

The communication between groups can take place in two ways: the information can propagate by means of links between groups or by passing through an intermediary user belonging to more than one group. We have defined as intermediary the links connecting a pair of users sharing a common group and with at least one of the users belonging also to a different group (see Fig.~1B). These users and their links have a high potential to pass information from one group to another in an efficient way~\cite{csermely06}. Several previous works pointed out to the existence of special users in Twitter regarding the communication in the network~\cite{asur11,wu11}. In order to estimate the efficiency of the different types of links as attractors of mentions and retweets, we measure a ratio $r_p^i$ for links in position $p$ and for interaction $i$ defined as
\begin{equation}
r_p^i = \frac{n_p^i}{N_p} ,
\end{equation}
where, as before, $n_p^i$ is the number of links with the interaction $i$ in position $p$ and $N_p$ is the total number of links in that position. The bar plot with the values of $r_p^i$ is displayed in Figure~5A. The efficiency of the different type of links can thus be compared for the attraction of mentions (red bars) and retweets (green bars). Links internal to the groups attract more mentions and less retweets than links between groups in agreement with the predictions of the strength of weak ties theory. Intermediary links attract mentions as likely as internal links: the fraction of intermediary links with mentions is very close to the fraction of internal links with mentions. This is expected because intermediary links are also internal to the groups. However, the aspect that differentiates more intermediary links from other type of links is the way that they attract retweets. Intermediary links bear retweets with a higher likelihood than either internal or between-groups connections (see Figure~5A and Section $1$ [Figs. S1-S4] in the Supplementary Information). This fact can be interpreted within the framework of the tradeoff between diversity and bandwidth~\cite{aral11}: strong ties are expected to be internal to the groups and to have high bandwidth, while ties connecting diverse environments or groups are more likely to propagate new information. High bandwidth links in our case correspond to those with multiple mentions, while links providing large diversity are the ones between groups. Intermediary links exhibit these two features: they are internal to the groups and statistically bear more mentions, and introduce diversity through the intermediary user membership in several groups. Although some theoretical works~\cite{granovetter73,aral11} suggest that ties with high bandwidth and high diversity should be scarce, we find that intermediary links are as abundant as internal links (see Fig.~2C). Moreover, in line with the theories~\cite{granovetter73,burt05,aral11}, higher diversity increases the chances for a link to bear retweets as can be seen in Figure~5B, which implies a more efficient information flow. In the inset of the Figure it is shown that the number of non-shared groups assigned to the users connected by the link positively correlates with a higher than expected number of retweets. 

\begin{figure*}
\begin{center}
\includegraphics[width=17cm]{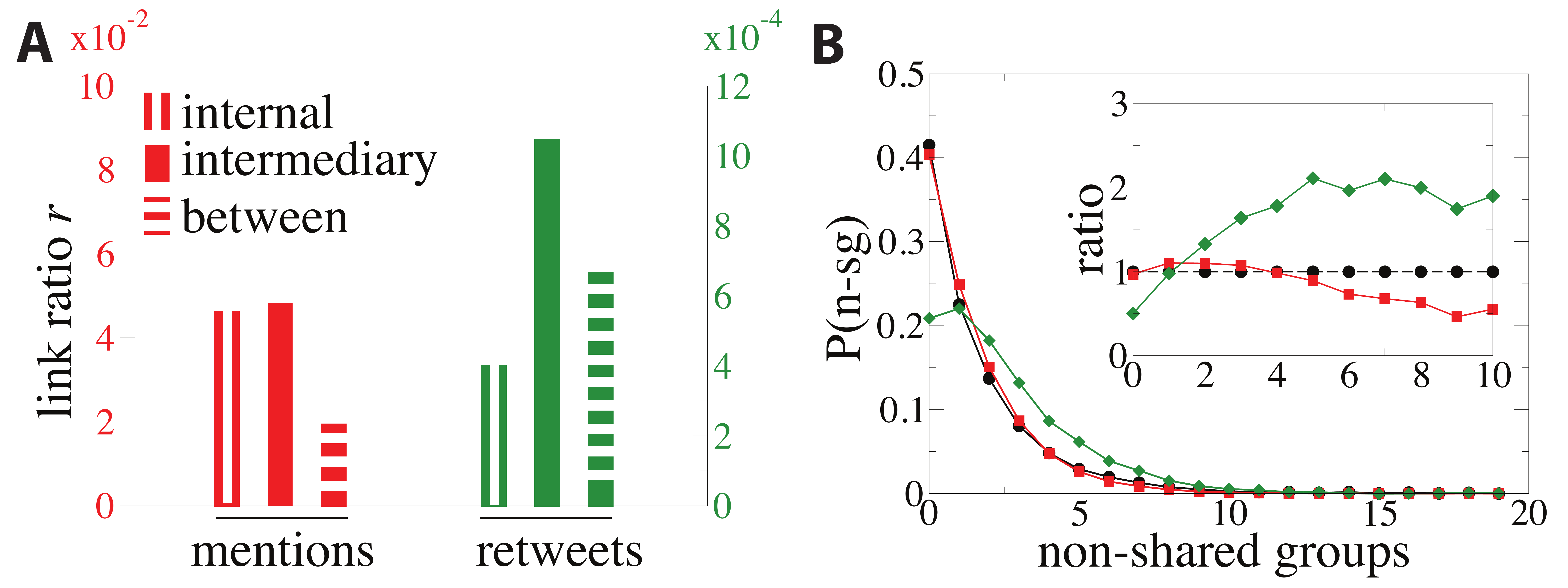}
\caption{ Intermediary links. (A) Ratio $r$  between the number of links with mentions or retweets and number of follower links. (B) Distribution of the links in the follower network (black curve), those with mentions (red curve) and retweets (green curve) as a function of the number of non-shared groups of the users connected by the link. Inset, ratios between these distributions and the follower network.}
\end{center}
\end{figure*}

\section{Discussion}

In summary, we have found groups of users analyzing the follower network of Twitter with clustering techniques. The activity in the network in terms of the messages called mentions and retweets clearly correlates with the landscape that the presence of the groups introduces in the network. Mentions, which are supposed to be more personal messages, tend to concentrate inside the groups or on links connecting close groups. This effect is stronger the larger the number of mentions exchanged and if they are reciprocated. Retweets, which are associated to information propagation events, appear with higher probability in links between groups, especially those that connect groups that do not show a high overlap, and more importantly on links connected to users who intermediate between groups. These intermediary users belong to multiple groups and play an important role in the spreading of information. They  acquire information in one group and launch retweets targeting the other groups of which they are members . At the same time, the access to new information can transform them into attractive targets to be retweeted by their followers. The relevance of certain users for the spread of information in online social media has been discussed in previous works. Our method provides a way to identify these special users as brokers of information between different groups using as only input the follower network.

From the sociological point of view, the way that the activity localizes with respect to the groups allow us to establish a parallelism with the organization of offline social networks.In particular, we have shown that the theory of the strength of weak ties proposed by Granovetter to characterize offline social network applies also to an online network. Furthermore, some of our results can be explained within the framework of Burt's brokerage and closure and Aral's diversity-bandwidth tradeoff theories. The specific properties of Twitter offers an opportunity to study directly the importance of the links for personal communications or for information diffusion. According to these theories, the strong social ties tend to appear at the interior of the groups or between close groups as happens for the links with mentions in Twitter. In addition, the socially weak ties are expected to be more common connecting different groups and to be important for the propagation of information in the network. This is similar to what we observe for the links with retweets that concentrate with high probability in links between dissimilar groups or in intermediary links. Besides the roles assigned by these two theories to the links, we have found that intermediary users and links are also an important component to take into account for understanding information propagation. These links tend to be characterized by high bandwidth and diversity in the context of Aral's study, and exhibit high information diffusion efficiency. Based on all these findings, despite the myth of one million friends and the doubts on the social validity of online links, the simplest connections of the online network bear valuable information on where higher quality interactions take place.

\section{Materials and methods}

\subsection{Description of the dataset}

\begin{table}[h!]
\begin{center}
\begin{tabular}{l | c | r |rrr}
\hline\hline
Property		&Follower		&Links with		&Links with	&\\
	         	&links	 		&mentions		&retweets	&\\[0.5ex]
\hline
Users	        &$2\,408\,534$	&$377\,760$		&$26\,480$	&\\
Links	        &$48\,776\,888$	&$1\,224\,484$	&$32\,169$	&\\
 Reciprocity 	&$27\%$		&$14\%$		&$0.7\%$	&\\
\hline\hline
\end{tabular}
\label{tab_data}
\end{center}
\caption{Overall characteristics of the follower network and of the interactions taking place on it.}
\end{table}

The data analyzed in this paper was collected in a two step process: the fist stage corresponds to the collection of the {\it follower network} (followers and followees), while the second consists in the retrieval of the user activity from the stream of Twitter (plain tweets, mentions and retweets). In the first stage, the directed unweighted network is obtained from the information on the followers and followees of each user. The data was collected using a breadth-first search technique: Starting from several seeds, followers and followees of the seeds were retrieved. Then the same procedure was repeated for the newly discovered users obtaining a so-called snowball sampling of the follower network. The procedure is stopped after several steps when the number of newly discovered users in $n$-th breadth is small compared with the total number of users already discovered in the $(n-1)$-th step. The process was run in November $2008$, gathering information for a total of $2\, 408 \, 534$ users. Due to the internal exploration of the network, one can anticipate that this method tends to detect the users with the highest in or out degree that belong to the largest connected cluster of the network.

The second stage consists in searching for all the tweets of the users found in the follower network for a period of time from November $20$ to December $11$. The activity dataset was constructed from these gathered tweets. The tweets containing usernames with a '@username' functional syntax were used for the mentions. Tweets that were reposted from other users, and which also hold a special format of the form 'RT @username', were used to build our retweet dataset. In some cases for mentions and retweets multiple users can be specified. Then we count only the first user for the purpose of our analysis. It is also worthy to note that mentions (replies) and retweets are now implemented into Twitter system~\cite{blog2}. The subset of retweets has been removed from a set of mentions to avoid overlap. In total, we obtained $12\, 486\, 784$ tweets from $587\, 142$ users in the network, what stands for $24\%$ of all users from the follower network. The rest of users either did not posted any tweet in their profile during the period of data collection ($80$-$90\%$ of cases), had a protected profile ($5$-$10\%$ of cases) or removed their profiles ($5$-$10\%$ of cases). Out of these tweets $1\, 742\, 956$ where mentions and $46\, 156$ where retweets. For the purpose of the analysis we have filtered out mentions and retweets which happened without underlying follower relation, in order to avoid inclusion of messages sent to not-known users and also to be able to perform comparisons with our baseline model consisting of the follower network. The resulting set of links with different interactions is summarized in Table~1. Note, that links with mentions/retweets can have multiple mentions/retweets happening over them.

The dataset is a good representation of what Twitter was at the end of $2008$ both in the social network and in the activity of the users. According to Ref.~\cite{blog1}, Twitter at the time of the data collection had less than $5$ million registered users. Therefore we estimate that our dataset contains information about more than $50\%$ of the most active users from that time. Other aspects of this dataset related to system scalability and trace generation were studied in Refs.~\cite{pujol10, pujol2, erramili}.

\subsection{The OSLOM clustering method}

OSLOM is a method based on a topological approach to detect statistically significant clusters~\cite{oslom11,significance10}. A null model that consists of graphs obtained by reshuffling the connections of the given network is considered. As a next step the probability of finding each group in the ensemble formed by these random graphs is estimated. During this procedure, it is assumed that an optimized clustering technique has been applied to the random graphs and therefore it is necessary to use techniques from the statistics of extremes and from order statistics to evaluate properly the probability of each group.  Oslom incorporates a local search method for the exploration of the network with the aim of finding clusters that improve the estimated probability, that is to find groups that have lower probability of existence in random graphs. OSLOM provides a set of clusters at the lowest hierarchical level and a list of nodes belonging to several groups and those not belonging to any group. The method has been tested in different benchmark networks containing planted groups, nodes belonging to several groups and nodes added to the network with random connections. Its high level of proficiency to recover the planted groups has been proved even when nodes with random connections are introduced in a graph with bona fide group structure. In those cases, OSLOM detects these nodes as no-group nodes~\cite{oslom11}.

\appendix

\section{Results with other clustering techniques}

\begin{center}
\begin{table*}
\caption{Summary of the results regarding internal connections when the groups are obtained with several clustering algorithms for different samples of the network. We measure the trend of the mentions to concentrate in internal connections. Legend: $w$ - weak signal, $sg$ - signal only for small groups, typically smaller than 10 members, a hyphen is inserted if we have no results. }
\begin{tabular}{lrr|rrrrrr|r}
\hline\hline
Network sample&Nodes&Edges&Oslom&Infomap&Moses&Real-time	&Louvain&Radatools&Figure \\
\hline
Whole network	&$2\,408\,534$	&$48\,776\,888$	&\tickYess	&\tickYess	&-		&\tickYess	&$w$		&-			&\ref{fig_panel_sel0}\\
Snowball	  2 hops 	&$61\,492$		&$5\,558\,036$	&\tickYess	&\tickYess	&\tickYess	&\tickNo		&\tickNo		&-		&  \ref{fig_panel_sel3nei2b}\\
Snowball	  3 hops 	&$175\,078$		&$10\,356\,020$	&\tickYess	&\tickYess	&\tickYess	&$w$			&\tickYess		&-	&\ref{fig_panel_sel3}\\
Random	200k	&$200\,000$		&$346\,578$		&\tickYess	&\tickYess	&\tickYess	&$sg$		&$sg$		&$sg$		& \ref{fig_panel_sel4}\\
No hubs			&$2\,395\,415$	&$23\,404\,103$	&-		&\tickYess	&\tickYess	&$w$			&\tickNo		&-		& \ref{fig_panel_sel6}\\
Oslom groups 	&$99\,832$		&$1\,216\,942$	&\tickYess	&\tickYess	&\tickYess	&\tickYess	&\tickYess	&\tickYess	&  \ref{fig_panel_sel2oslom5k}\\
\hline
\end{tabular}
\label{tab_clus}
\end{table*}
\end{center}

In this section, we check the reliability of the  results of the main text regarding the localization of the activity when the groups are obtained with clustering techniques different from Oslom. The reasons to select Oslom as the main method are  that (i) the software is publicly available, (ii) the method is able to analyze the full directed follower network in a reasonable amount of time, (iii) it detects the overlapping communities (bridging nodes and bridges) and nodes not belonging to any group and (iv) the clusters obtained are statistically significant according to a clear null model ~\cite{oslom11,oslom2}. We do not ask the same properties to other methods explored here but they should meet the following conditions: the methods should be available online in the form of software tools, they should be able to deal with relatively large samples of dense graphs, and, if possible, they should include a version to analyze directed networks. We have found several methods satisfying fully or partially these conditions and so we will show in the remainder results with groups detected in the follower network (or in sub-sampling of it) by Infomap~\cite{rosvall08,rosvall2,rosvall3}, Moses~\cite{mcdaid10,moses2}, a message-passing algorithm proposed by Raghavan et al~\cite{raghavan,leung} that we will refer to as Real-time community detection and two algorithms to optimize modularity~\cite{newman}: the Louvain method for community detection~\cite{louvain,blondel2} and a slower  modularity optimization algorithm adapted to deal with directed networks that was implemented in the software Radatools~\cite{alex2,radatam}.

\begin{figure}
\begin{center} 
\includegraphics[height=8.6cm]{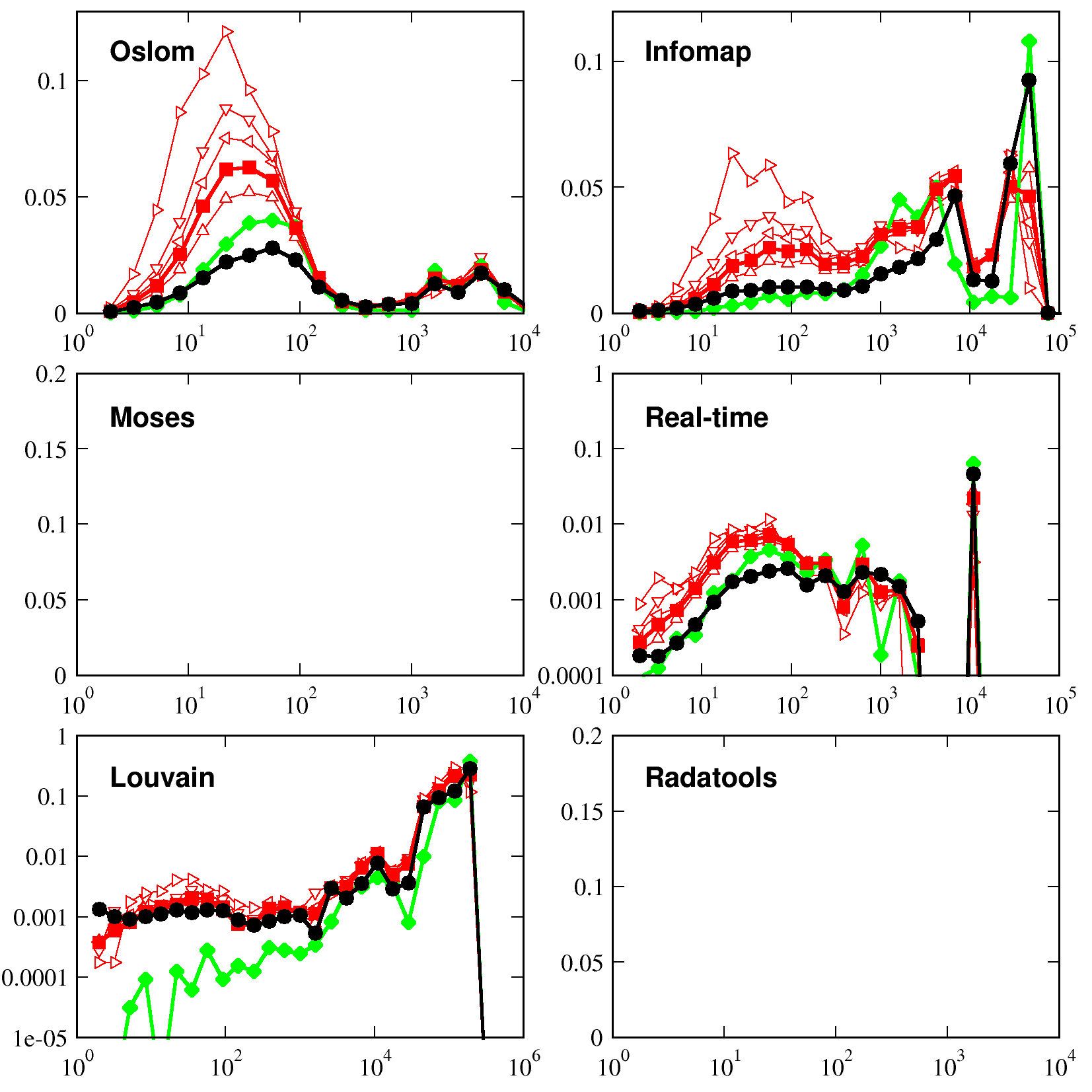}
\caption{Internal activity for different clustering algorithms from left up corner to the right: Oslom, Infomap, Moses, Louvain, Real-time community detection, and Radatools. Fraction of links of different types internal to the groups as a function of the group size in number of users. The black curve is for the follower network, which acts as baseline for the links with any mentions (red curve with closed square symbols) and for links with specific number of mentions (red curves with open triangle symbols rotated 90 degrees counterclockwise starting from straight up triangle: 1 mention non-reciprocated, 3 mentions, 6 mentions, and more than 6 mentions reciprocated).}
\label{fig_panel_sel0}\end{center}
\end{figure}

\subsection{Description of the clustering methods}

Cluster detection in graphs is a topic of very active research~\cite{santo10}. There are plenty of methods in the literature based on different techniques and with different features and capabilities. The methods selected here are representative of some of the most popular approaches used today for searching groups in networks. Modularity optimization is based on a comparison  between the number of internal links and the average expected number in a random graph. It has been one of the most popular community detection methods in the last few years, although it is not free problems such as resolution limits~\cite{santo} or difficulties to find the absolute maximum of the modularity due to a rough landscape of its value in the space of the possible network partitions~\cite{clauset}. The Louvain method is based on a contraction of the network similar to real-space renormalization and attempts to keep the modularity function constant at each contraction step. Moses includes an overlapping stochastic blockmodelling approach as the basis to its community detection algorithm. It provides a procedure for the optimization of the log-likelihood introduced by the blockmodelling approach. Oslom is based on a structural approach comparing the internal and external links with the best expectations in an equivalent random graph. In this method, the fact that clustering algorithms follow a process of optimization is incorporated to give an assessment of the probability of finding a similar group in a random network. The method then searches for groups in the given network that have low probability of appearance in the equivalent random graphs. The last tested method, Infomap, is based on a different approach, trying to optimize the information fluxes in the given graph by the compartmentalization of the network. Infomap is based on the optimization of the description of random walkers paths in the network using information theory concepts

\begin{figure}
\includegraphics[height=8.6cm]{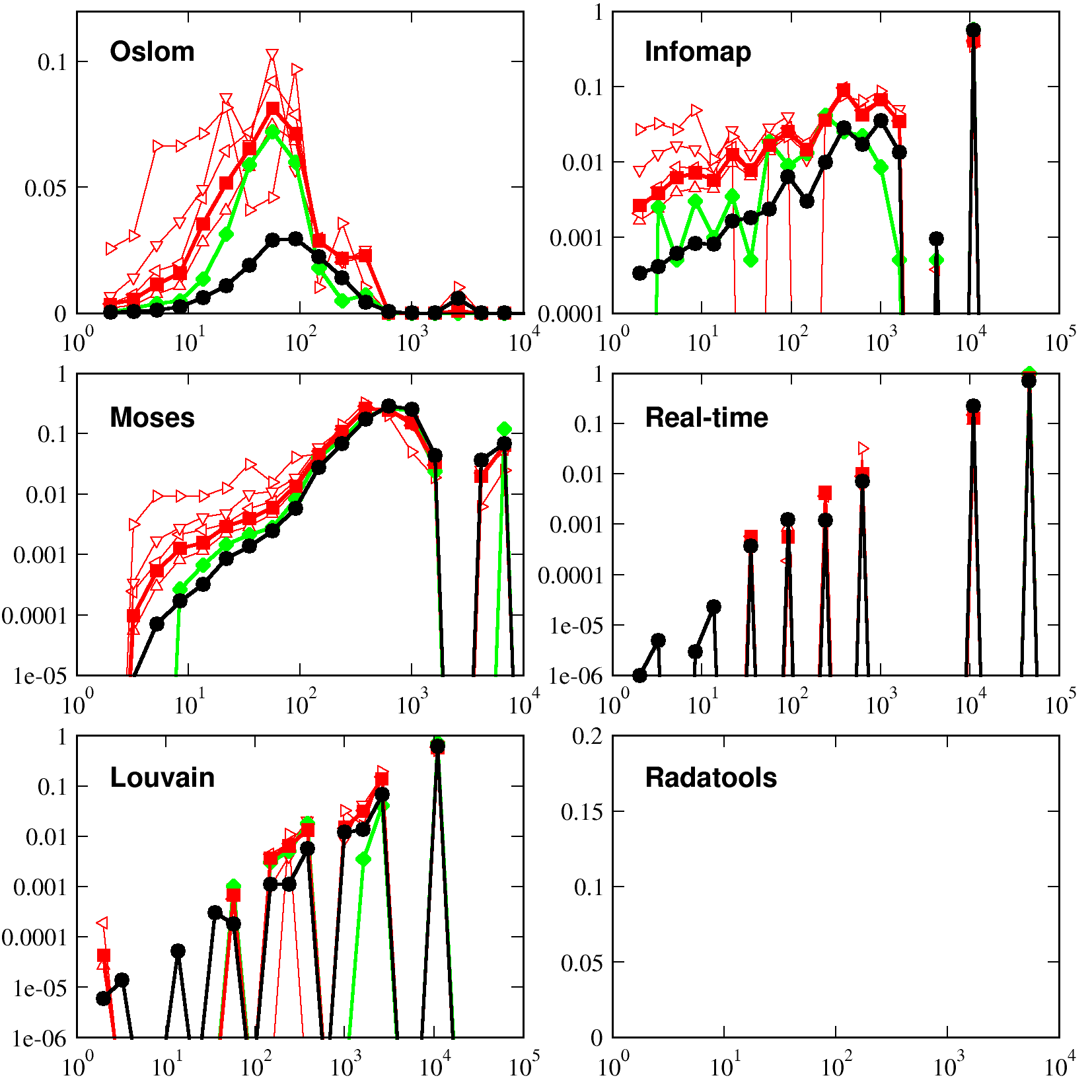}
\caption{Internal activity for different clustering algorithms run for the snowball sample of the network (2 neibhors away from a random seed), from left up corner to the right: Oslom, Infomap, Moses, Louvain, Real-time community detection, and Radatools.}
\label{fig_panel_sel3nei2b}
\end{figure}

\begin{figure}
\includegraphics[height=8.6cm]{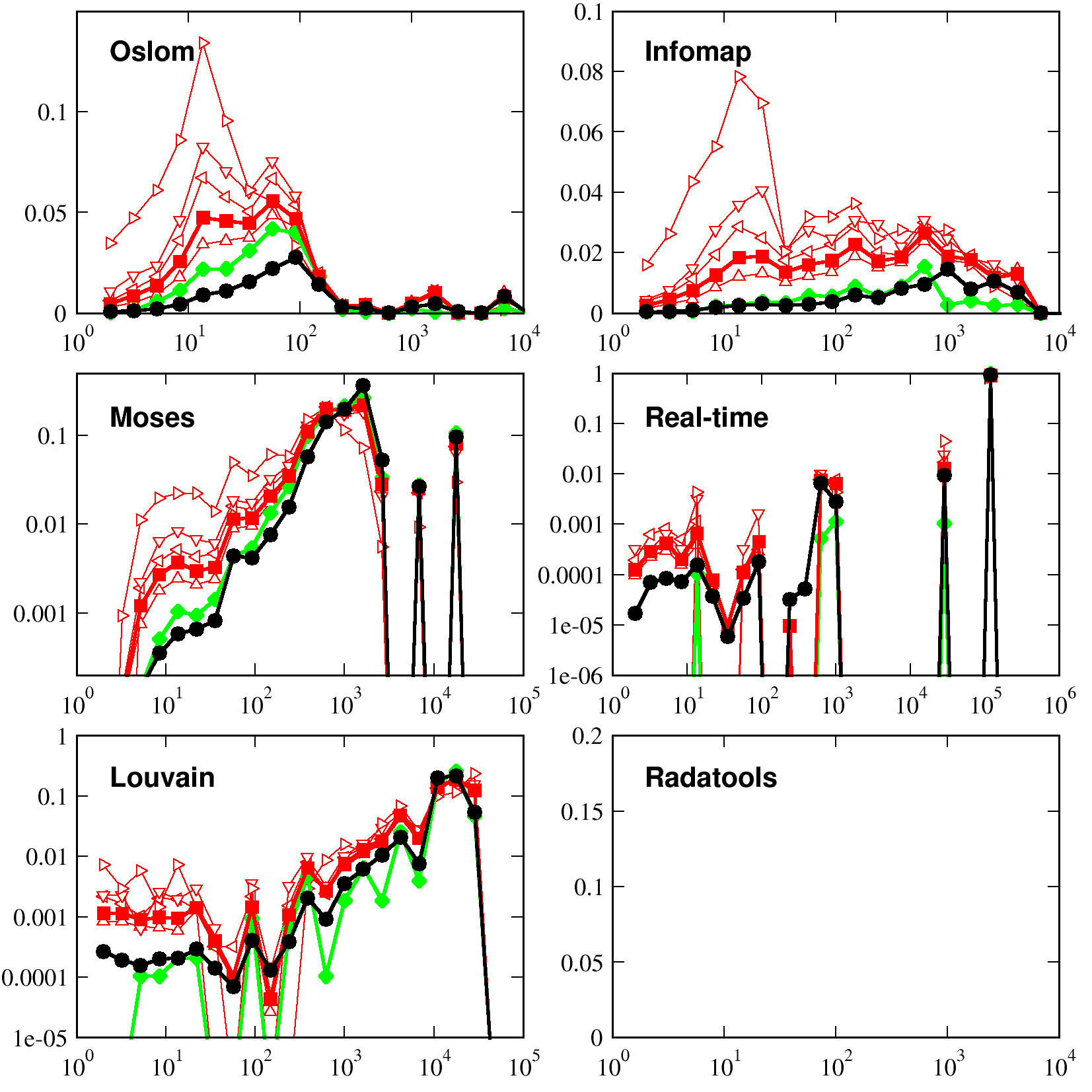}
\caption{Internal activity for different clustering algorithms run for the snowball sample of the network (3 neighbors away from a random seed), from left up corner to the right: Oslom, Infomap, Moses, Louvain, Real-time community detection, and Radatools.}
\label{fig_panel_sel3}
\end{figure}

The clustering methods can be distinguished by other aspects apart from the approach that they use to find groups. To start with, some as Infomap, Radatools as well as Oslom have specific versions to analyze directed graphs. We used them on the original or sub-samples of the directed follower network. In order to use the other three methods (Moses, Louvain, Real-time), we have symmetrized the network. The symmetrization consists in ignoring the directionality of the links, and considering them as undirected. This procedure neglects information that can be important to define the groups and can affect the performance of the methods. A second difference is the ability to find overlapping communities or bridging nodes. Only Oslom and Moses are able to detect users belonging to more than one group. And, finally, the performance of the methods varies. A way to compare clustering methods is to generate benchmark networks in which the groups are a priori known, then to increase the level of disorder in the connections and to test up to which point the methods recover the planted groups. Infomap was found to be one of the best performing methods in this sense in a recent comparative work~\cite{lanci_comp}, while Oslom has been thoroughly tested in a later work~\cite{oslom11} getting results that are comparable to Infomap's.

\begin{figure}
\includegraphics[height=8.6cm]{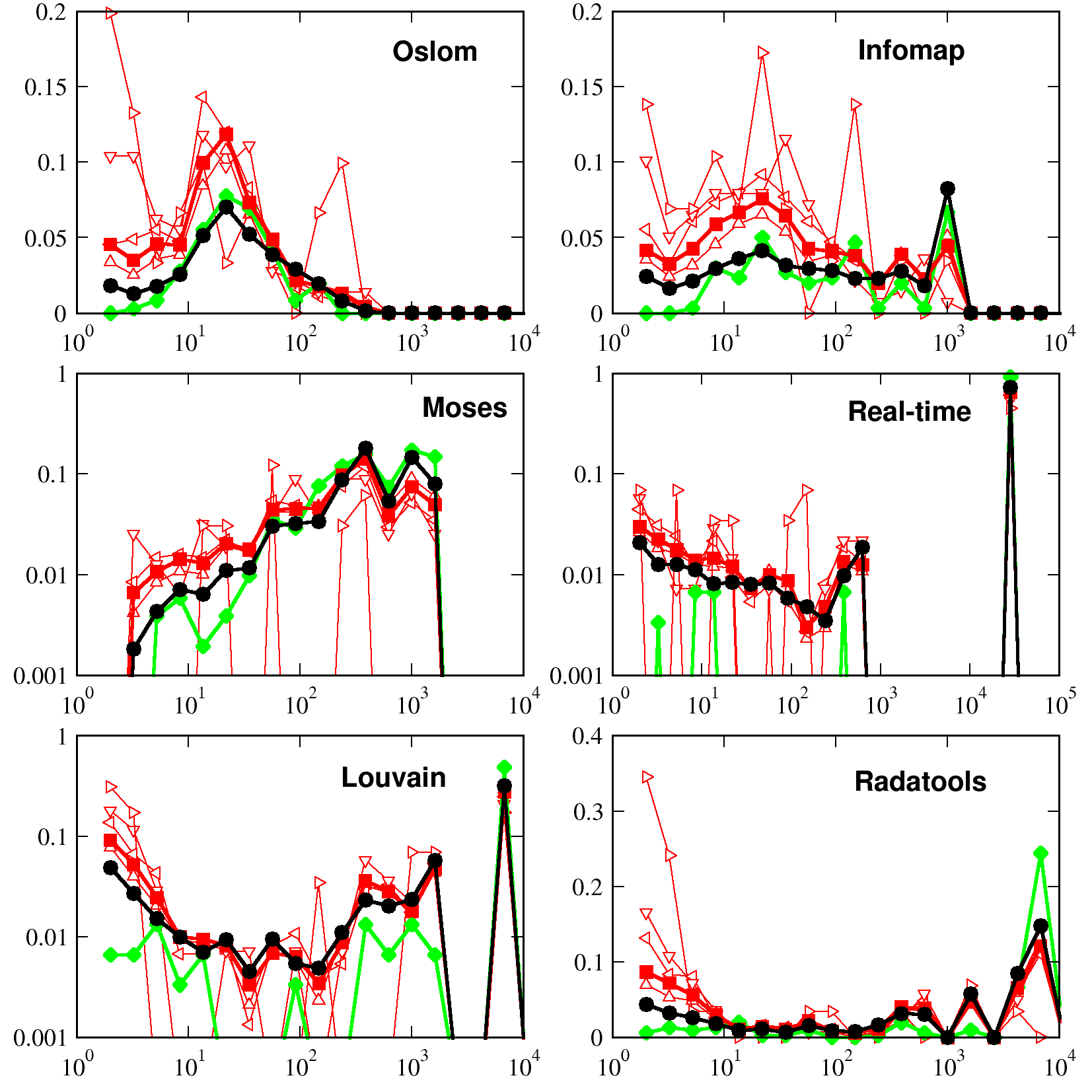}
\caption{Internal activity for different clustering algorithms run for the subgraph of randomly chosen 200k nodes, from left up corner to the right: Oslom, Infomap, Moses, Louvain, Real-time community detection, and Radatools.}
\label{fig_panel_sel4}
\end{figure}

\begin{figure}
\includegraphics[height=8.6cm]{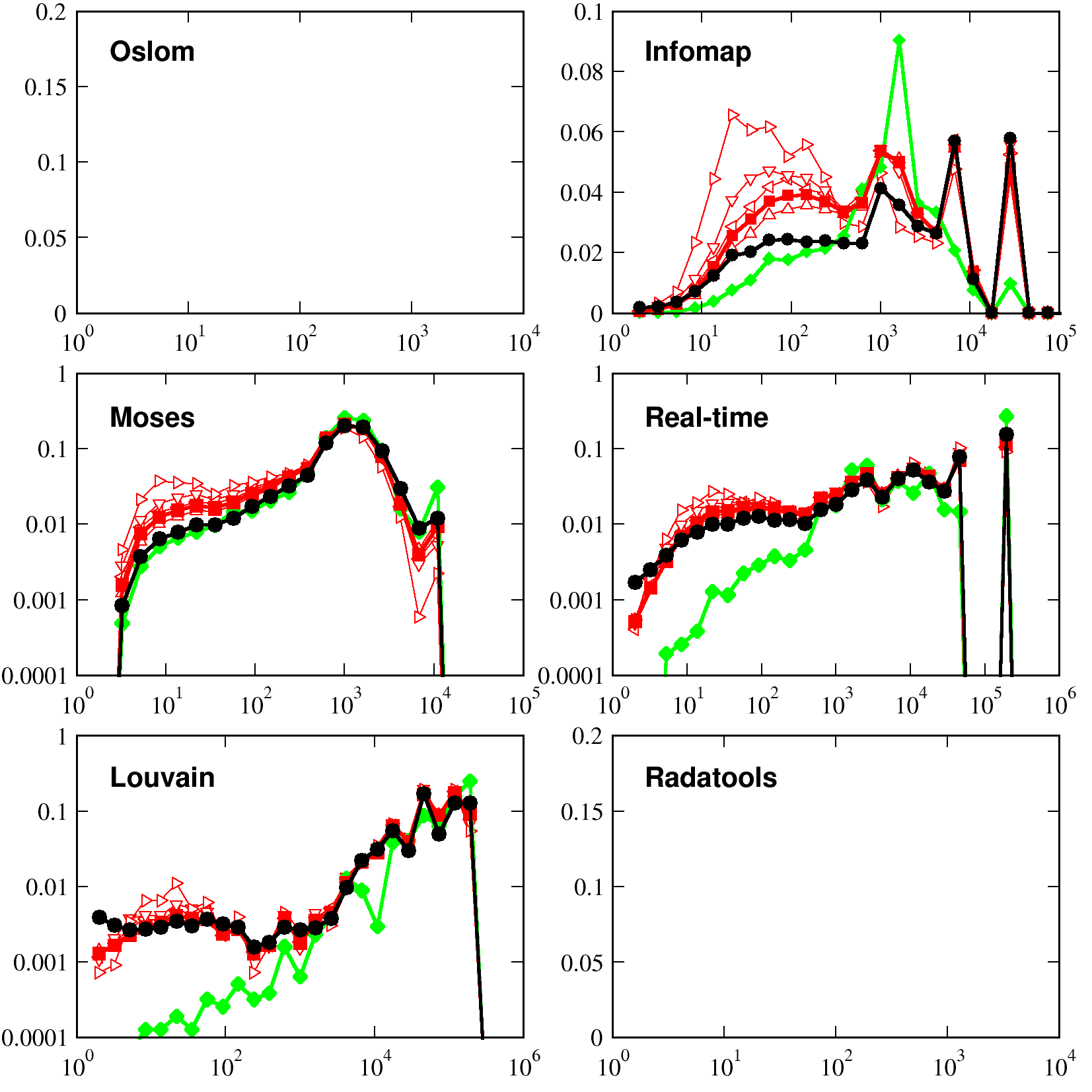}
\caption{Internal activity for different clustering algorithms run for the network with removed hubs, from left up corner to the right: Oslom, Infomap, Moses, Louvain, Real-time community detection, and Radatools.}
\label{fig_panel_sel6}
\end{figure}

\begin{figure}
\includegraphics[height=8.6cm]{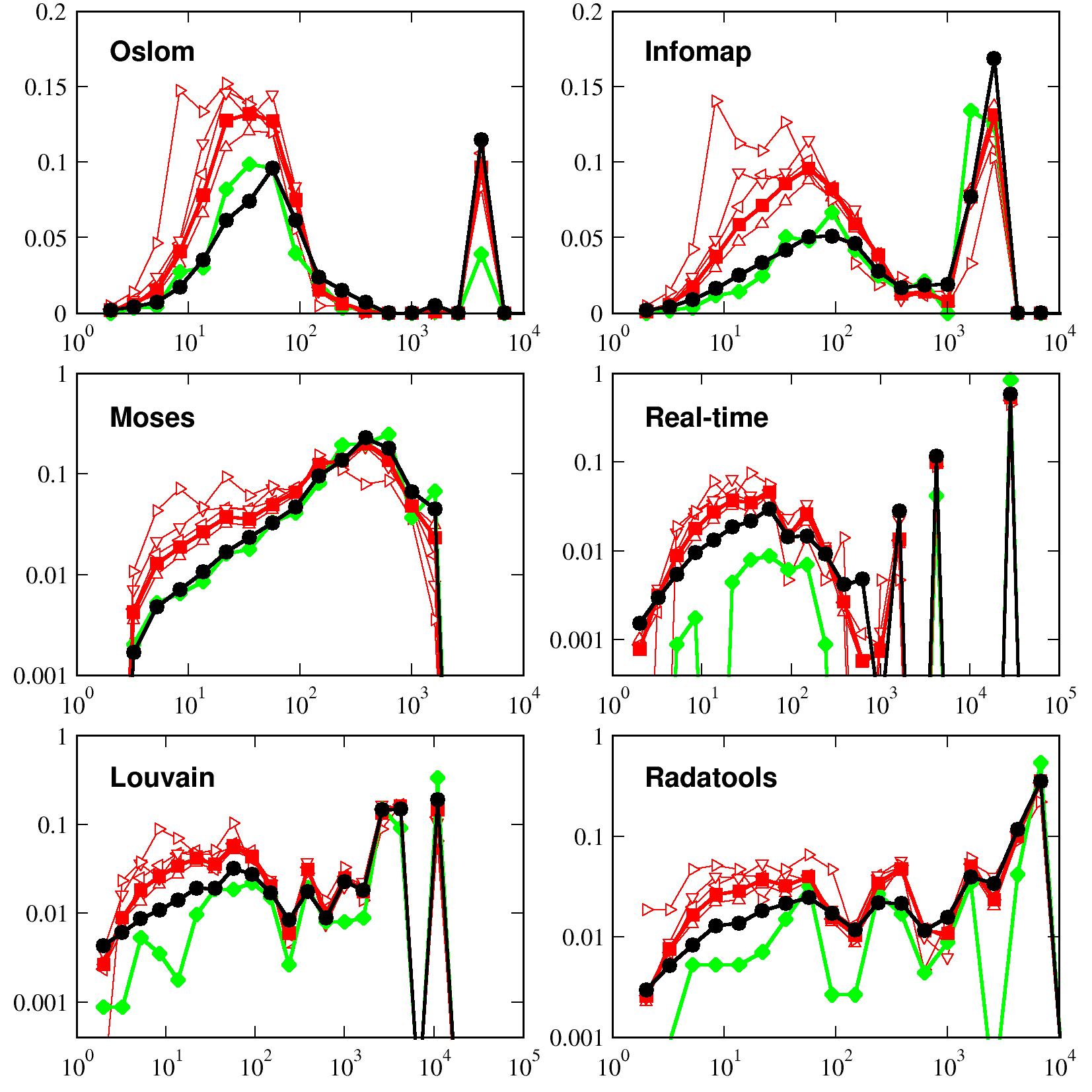}
\caption{Internal activity for different clustering algorithms run for the subgraph build from 5000 randomly selected groups found by Oslom, from left up corner to the right: Oslom, Infomap, Moses, Louvain, Real-time community detection, and Radatools.}
\label{fig_panel_sel2oslom5k}
\end{figure}

A final point to consider is that not all the methods are able to cope with the large size and the high density of our empirical follower network. The computational cost of the analysis makes difficult for some of them to deal with large networks. For this reason, and as can be seen in Table~\ref{tab_clus}, we have run the methods against several samplings of the follower network. These samplings include, when possible, the full network and, when not, subnetworks extracted with different procedures. We have  implemented a snow-balling technique similar to the one that led to the collection of the original network and obtained subgraphs using shells of $2$ and $3$ hops starting from different initial seeding nodes. We have also generated subgraphs by choosing $200\,000$ nodes at random and extracting all the links between them. Another network sample has been extracted by considering the full graph without the hubs (nodes with total degree higher than $1000$). This idea comes from the fact that in Twitter there can be users such as celebrities with thousands and millions of followers but whose connections bear no information on bona fide social activity in the network. And, finally, as a sanity check, we have also generated a sub-graph formed by some of the groups detected by Oslom. 

\subsection{Internal links}

In the results of the main text, the links with mentions are more abundant than the baseline inside groups of size up to $150$ users. Larger groups do not seem to behave in the same way and the abundance of links with mentions fall upon the baseline. We find a similar signal when the groups are extracted with other clustering algorithms, see the summary of Table~\ref{tab_clus}. The results for the  full network for $2$ out of $3$ algorithms tested are in qualitative correspondence with the results of Oslom. These include Infomap's results for all the network samples (Infomap is supposed to be one of the most trustable methods for community detection~\cite{lanci_comp}). The results, together with results of Oslom for the same sample of the network, are depicted in Fig. \ref{fig_panel_sel0}. The figure reveals that the fraction of links with mentions inside of groups is higher than the fraction of any links inside groups irrespectively of the algorithm used. In case of all clustering algorithms (including Oslom), for both the sample and the full network, the effect is not visible for groups larger than $150$-$5000$ users (this number varies for different algorithms). When we take into account the number of mentions and whether they are reciprocated, the results show a remarkably consistent pattern. The more mentions, especially reciprocated, the link has, the higher the probability that it is inside of a small group independently of the community detection algorithm used or the sample of the network considered (see Figs.~\ref{fig_panel_sel3nei2b} to \ref{fig_panel_sel2oslom5k}).

\begin{figure}
\begin{center}
\includegraphics[width=8.6cm]{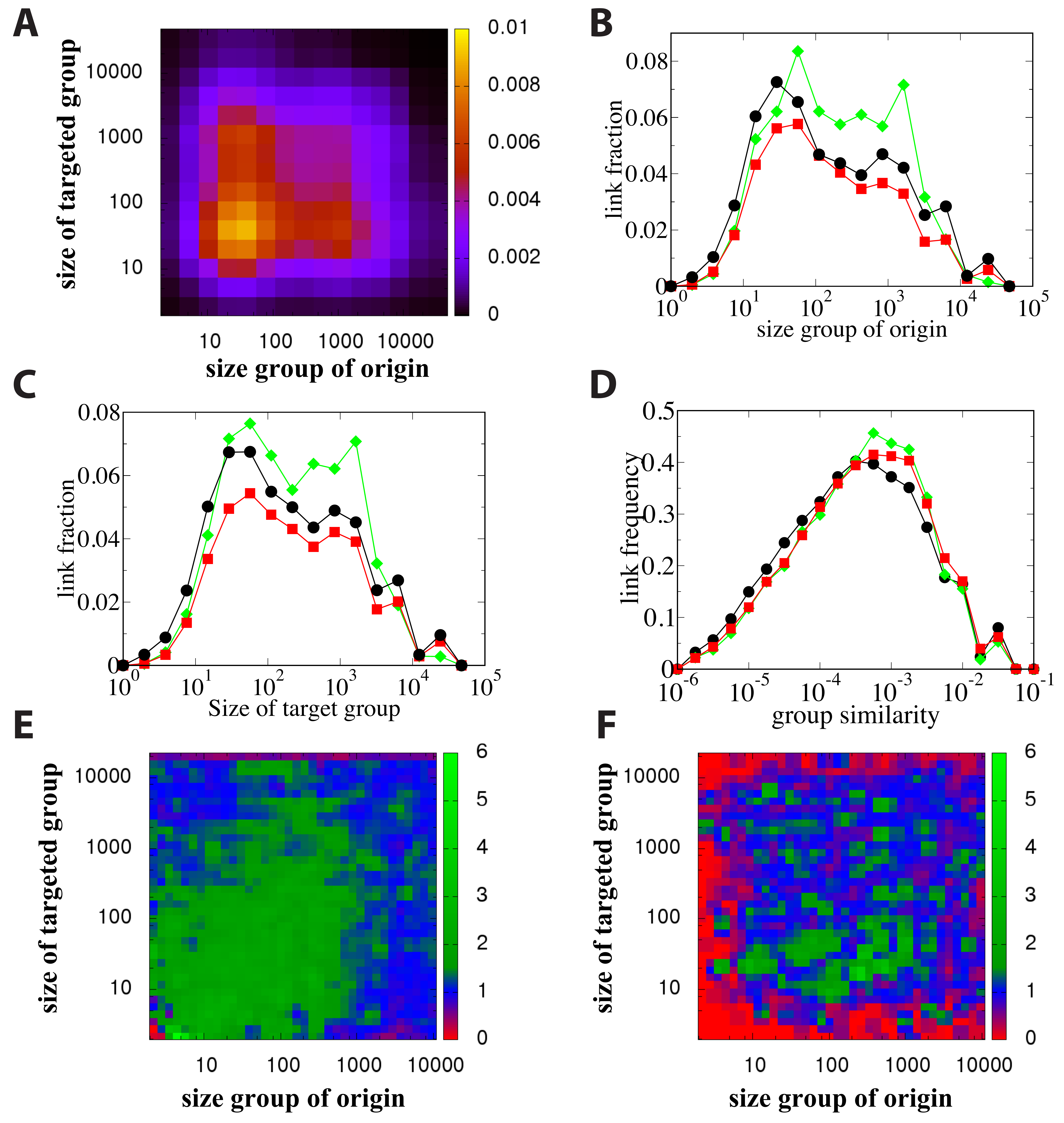} 
\caption{Activity on between-groups links when the groups are detected by Infomap in the sample without hubs. The panel reproduces the structure of Figure 4 of the main text and of Figure 17. (A) Fraction of links in the follower networks as a function of the size of the group of origin and destination. (B) and (C) Fraction of links of different types: follower relations (black circles), links with mentions (red squares) or with retweets (green diamonds), as a function of the size of the group of origin or destination, respectively. (D) Frequency of links of the different types as a function of the group-group similarity. Ratio between the average group similarity for the links between groups with mentions (E) or retweets (F) and the follower network as function of the size of the group of origin and destination.}
\label{fig_infomap}\end{center}
\end{figure}

\subsection{Between-groups links}

In order to check whether the results on the localization patterns of the activity in the links between groups discussed in the Figure 4 of the main text and in Figs~\ref{maps} of this Supplementary can be reproduced with groups obtained by other clustering methods, we have repeated the analysis of the group-group links using the groups found by Infomap. The results are in Figure~\ref{fig_infomap}. 
Even though the shape of some of the curves is different, the main qualitative results confirm the trends observed with Oslom regarding the concentrations of mentions in links between more similar groups, and retweets in those connected groups with medium or low similarity. Also the between-group links concentrate a higher quantity of retweets and slightly less mentions.

\begin{figure}\begin{center}
\includegraphics[width=8.6cm]{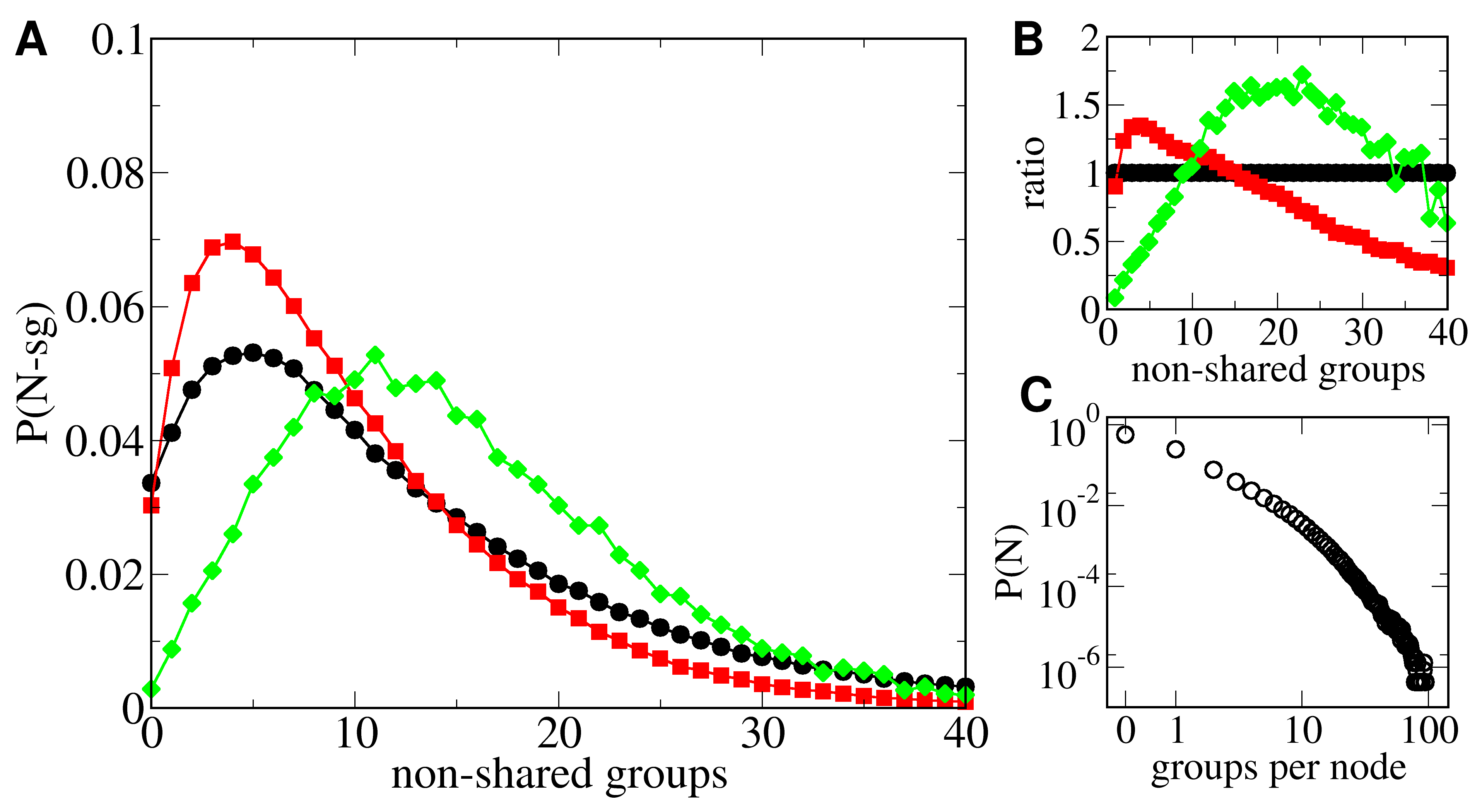}
\caption{Bridges between groups detected by Moses for the network sample without hubs. (A) Distribution of the links in the follower network (black curve), those with mentions (red curve) and retweets (green curve) as a function of the number of not-shared groups of the users at the extreme of the link. (B) Ratio between these distributions taking the follower network as baseline. (C) Distribution of the number of groups to which each user is assigned.}
\label{fig_moses}\end{center}
\end{figure}

\subsection{Bridges}

The role of bridging users and bridging connections can be investigated with clustering algorithms capable of detecting overlapping communities, and so capable of assigning nodes to more than one group (Moses and Oslom). The results for Oslom has been presented in the main text. Here in Fig.~\ref{fig_moses}A we show  results for the Moses algorithm, run for the sample of the network with removed hubs. The shape of the curves, especially for ratios between distributions for different types of links shown in Fig.~\ref{fig_moses}B, is consistent for the two clustering algorithms. The probability of having a retweet over a bridging link, which is proportional to the ratio, steadily grows from almost zero with the number of non-shared groups of the pair of interacting users until it reaches a maximum around value 20, and then it drops down. The pattern for links with mentions remains consistent for both algorithms, with small maximum for smallest values of number of non-shared groups and steady decay for larger values.

\subsection{Alternative procedures to validate our results}

\begin{figure}
\begin{center}
\includegraphics[width=8.6cm]{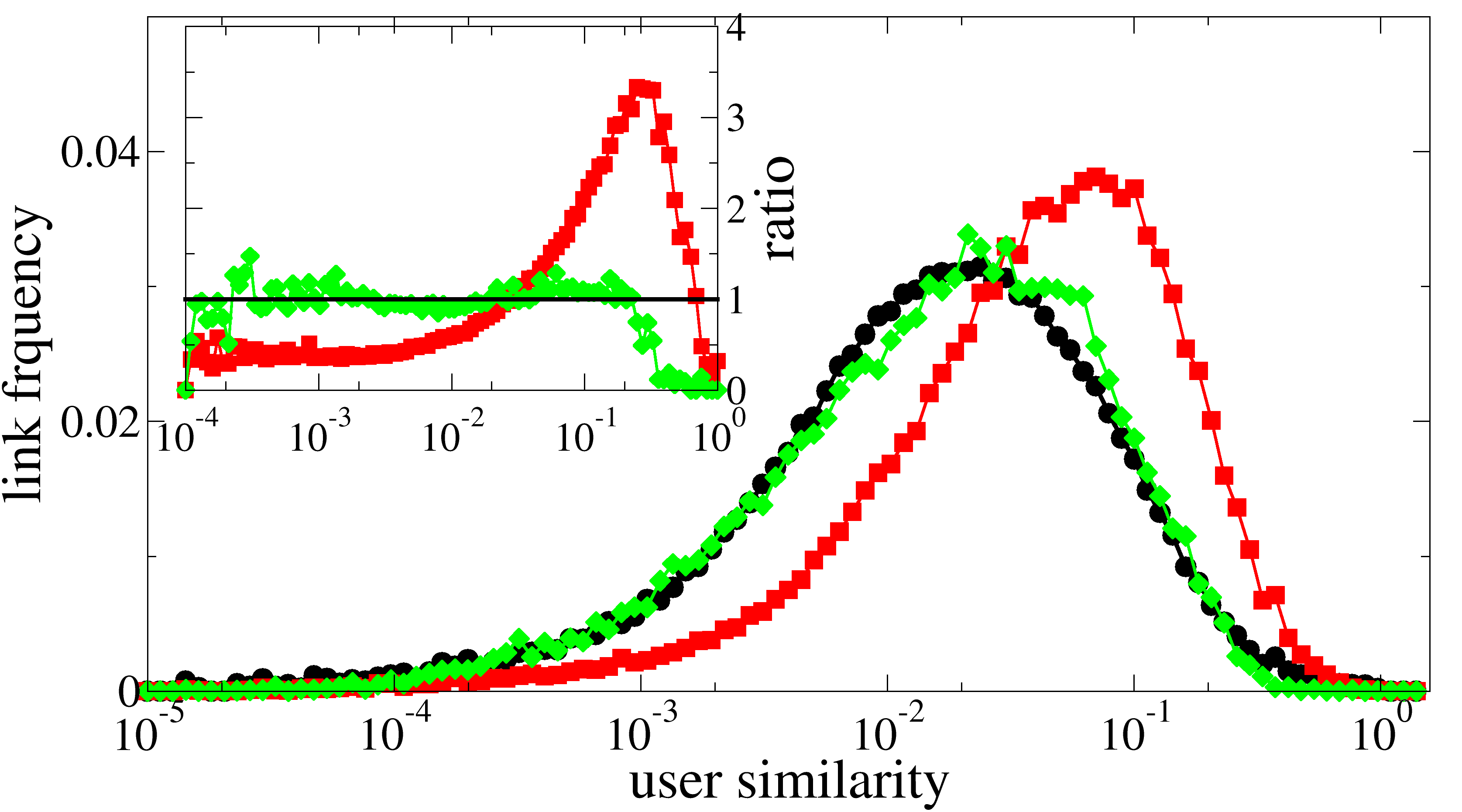}
\caption{Jaccard similarity of users followers. Users similarity frequency for pairs of users connected by a follower link (black circles), by a link with a mention (red squares) and a link with retweet (green diamonds). Inset: ratio between these frequencies taking the follower network as a baseline. }
\label{fig_usim}\end{center}
\end{figure}

The theory of strength of weak ties have two formulations. The macroscopic formulation predicts strong ties to be inside of communities, whereas weak ties as the connectors between these communities. The microscopic formulation states that strong ties happen between users having many friends in common. In the paper, we focus on the communities but here we also show that in fact the microscopic predictions can be confirmed in Twitter as well. Jaccard similarity of two users is equal to number of shared followers by the two users divided by total number of unique followers the two users have. In Fig. \ref{fig_usim} we plot the distribution of the similarity for pairs of users who either send a mention or a retweet to each other or are simply connected by a follower tie. The distribution of similarity for the interaction which is supposed to be the most personal (link with mention) is shifted to the right, showing that indeed mentions happens much more often between users who share friends. This result and the finding, that mentions are more abundant inside of the communities, are consistent with the expectations of the theory of the strength of weak ties.

\section{Balance between the number of internal links and links between groups}

\begin{figure} \begin{center}
\includegraphics[width=8cm]{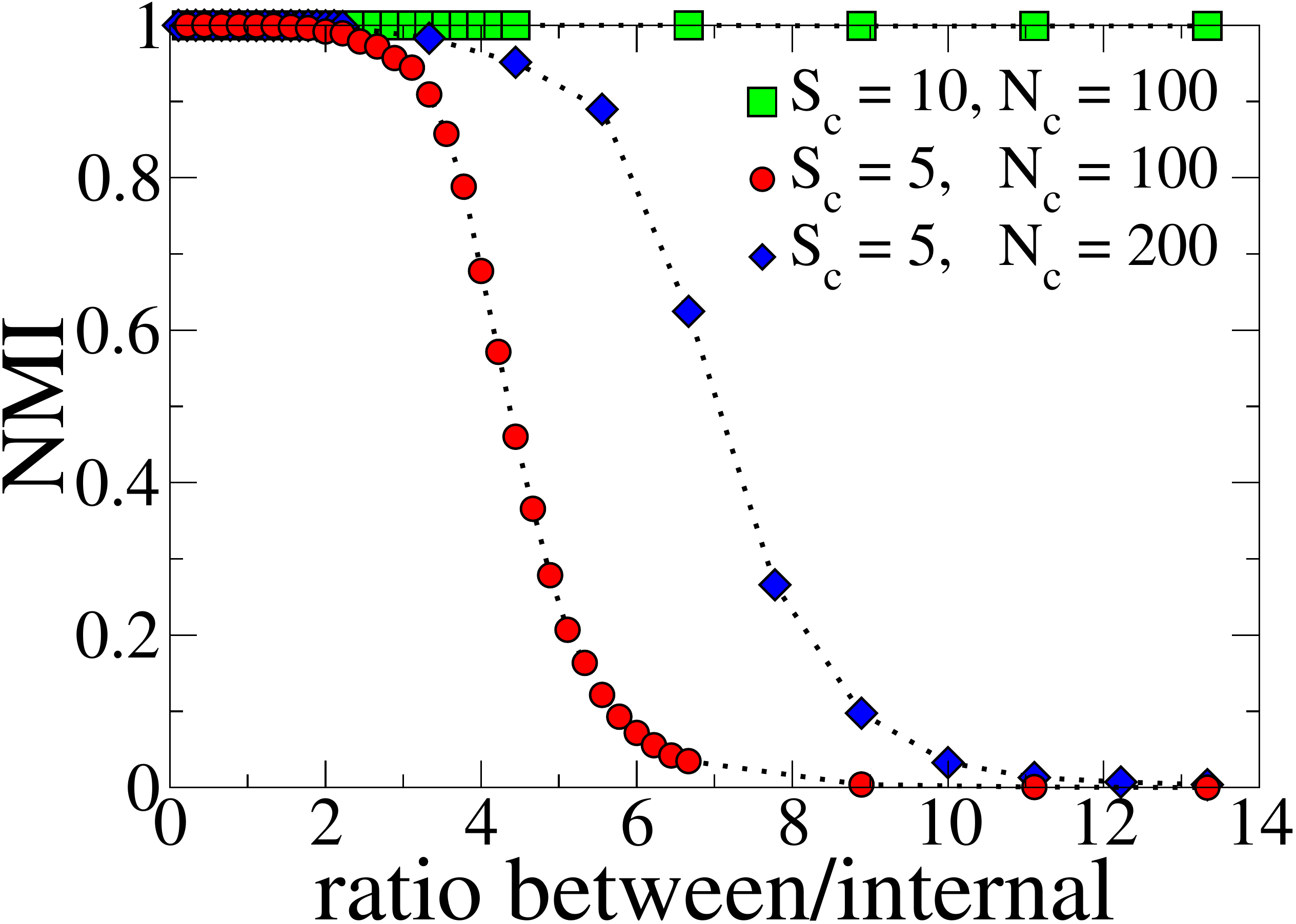}
\end{center}\caption{Normalized mutual information as a function of the ratio between the number of links between groups and internal links to the groups in a benchmark. The benchmark is composed of $N_c$ cliques (fully connected subgraphs) of size $S_c $ each. }
\label{nmi}
\end{figure}

In this section, we discuss in more detail the imbalance between the number of internal and between-group links that is seen in Figure 3C and the effect it may have on cluster detection. The objective of a clustering algorithm is to find areas of the network relatively isolated and dense in internal connections. How is then possible that the number of overall internal links is lower than that of links between groups in the clusters found by Oslom? The answer is that Oslom (as many community detection algorithms) is not attempting to optimize the balance between internal and between-group links in a direct way. The method searches for areas denser in internal connections than a baseline established by the properties of the random  graphs obtained by reshuffling the links of the original network while maintaining the nodes' degrees constant~\cite{molloy,oslom11}. To illustrate this idea, we have generated a benchmark formed by $N_c$ cliques (fully connected subgraphs) of size $S_c$ each. The final graph is then obtained by the addition of  $L_{bet}$ links between groups connecting nodes of different cliques at random. To quantify the level of similarity between the original cliques and the groups detected by Oslom, we use the normalized mutual information between partitions $NMI$ \cite{alex,lanci}. This quantity is equal to one when the two divisions of the network in groups--the original cliques and the groups detected-- are strictly equal and tends to zero when there is no relation between them. The results as function of the ratio between the number of links between groups and the number of internal links ($L_{int} = N_c \, Sc \, (S_c -1)/2$) is shown in Figure ~\ref{nmi}.  Oslom is able to detect the planted cliques up to high levels of the ratio internal vs between links, higher in any terms than the values seen in the real follower network. At the same time, the performance of the method improves with larger groups and with a larger number of cliques.

The reason for this ability is that the connections between groups are introduced at random, without any clear statistical preference for connections between two particular groups. Oslom can detect these random links and ignore them to evaluate which nodes belong to each group despite the high ratio of between-groups links over internal links.

\section{Statistical features of the groups detected with Oslom}

\begin{figure}
\begin{center}
\includegraphics[width=8.6cm]{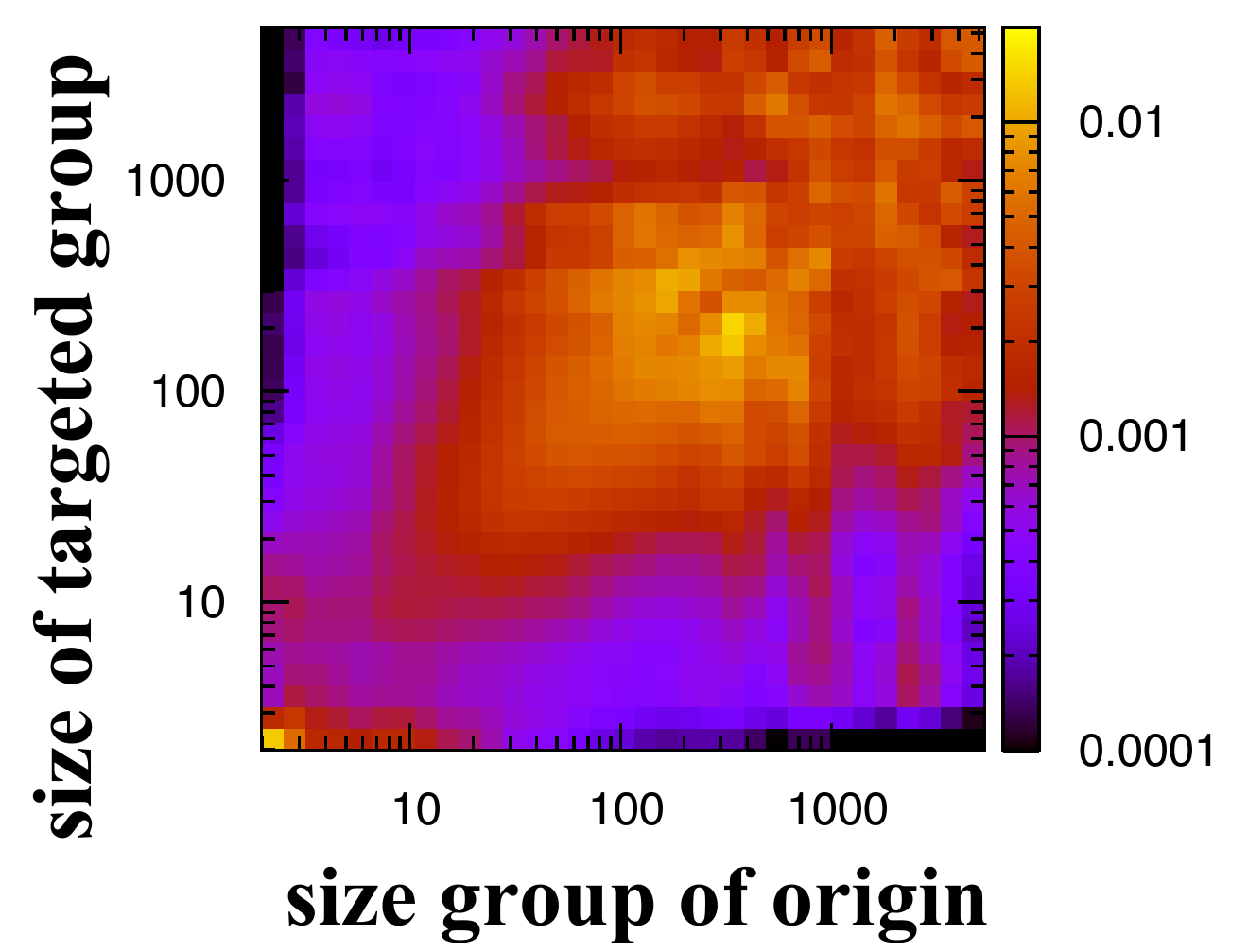}
\caption{Averaged group-group similarity for groups paired by follower links as a function of the groups sizes.}
\label{ggsim}\end{center}
\end{figure}

\begin{figure}
\begin{center}
\includegraphics[width=8.6cm]{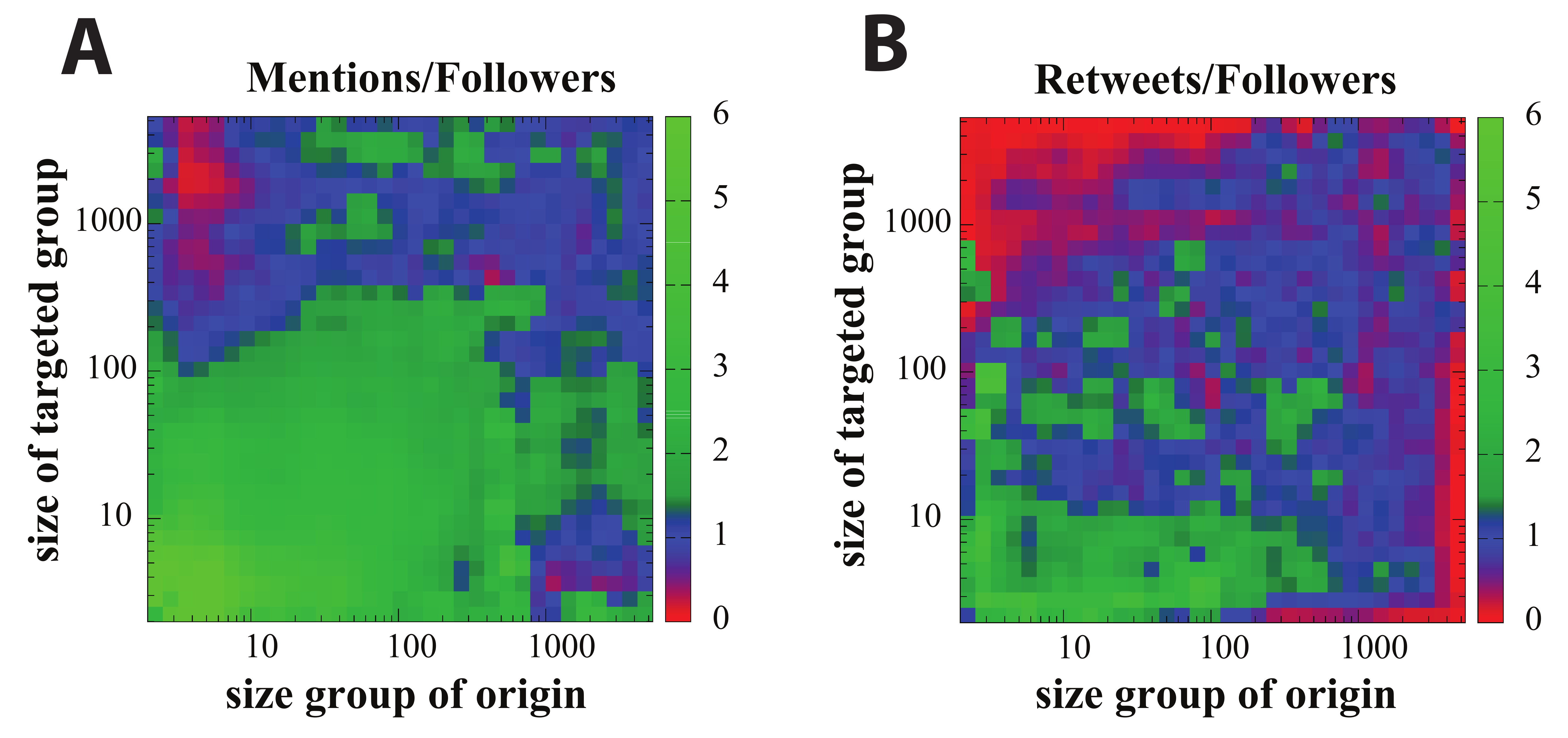}
\caption{Ratio between the average group similarity for the between-group links with mentions (A) or retweets (B) and the follower network as function of the size of the group of origin and destination.}
\label{maps}\end{center}
\end{figure}

We revisit the group similarity as defined in the main text. It is important to recall that the group-group similarity is defined as the Jaccard index of the connections of two groups (the ratio between the number of links in common and total links). We consider as the connections of a group all links originating from or targeting the nodes of the group. The similarity of the groups is dependent on the size of the groups of origin and destination of the particular link. Figure~\ref{ggsim} shows how the average similarity of groups connected by follower relations depends on the size. The largest similarity concentrates in links connecting groups of similar size. Although one should keep in mind the Figure 4A of the main text that shows where most of links are located, namely between groups of size from $10$ to $200$, making this region the most relevant. Taking the average similarity of the groups for the links with mentions and for the links with retweets and plotting the ratio of these quantities divided by the baseline average similarity of the follower network links, we obtain Figures~\ref{maps}A and \ref{maps}B. The signal reproduces the results of the overall histogram in the Figure 4D of the main text: (i) The links with mentions tend to connect groups with higher similarity compared to those with retweets or the baseline given by the follower network and (ii) The retweets normally happen in links between groups with a medium value of the similarity. The same picture is observed almost independently of the size of the group of origin and destination except for some areas that anyway concentrate a lower density of links.

Finally, we show results that complement the discussion on bridges of Figure 5 of the main text. The fraction of follower links that are bridges, bridges with mentions or with retweets as a function of the groups size can be seen in Figure~18A. Note that although a node can belong to more than one group, the links usually belong to a single group unless they  connect two bridging nodes. The size of the group in Figure~18A refers to each of the groups a bridging link belongs to. If a link connects two bridging nodes and so belongs to several groups, the link is counted as many times as groups it is in. As mentioned in the main text, bridges are very effective in attracting retweets. This behavior, present also in Figure~18A, contrasts with the one observed in Figure 3 of the main text for internal links. However, what is similar between internal and bridges is their attraction for mentions. This attraction increases with the number of times a link has been used for mentions (see Figure~18B), also in parallel to the results for pure internal links.

\begin{figure}
\begin{center}
\includegraphics[width=8.6cm]{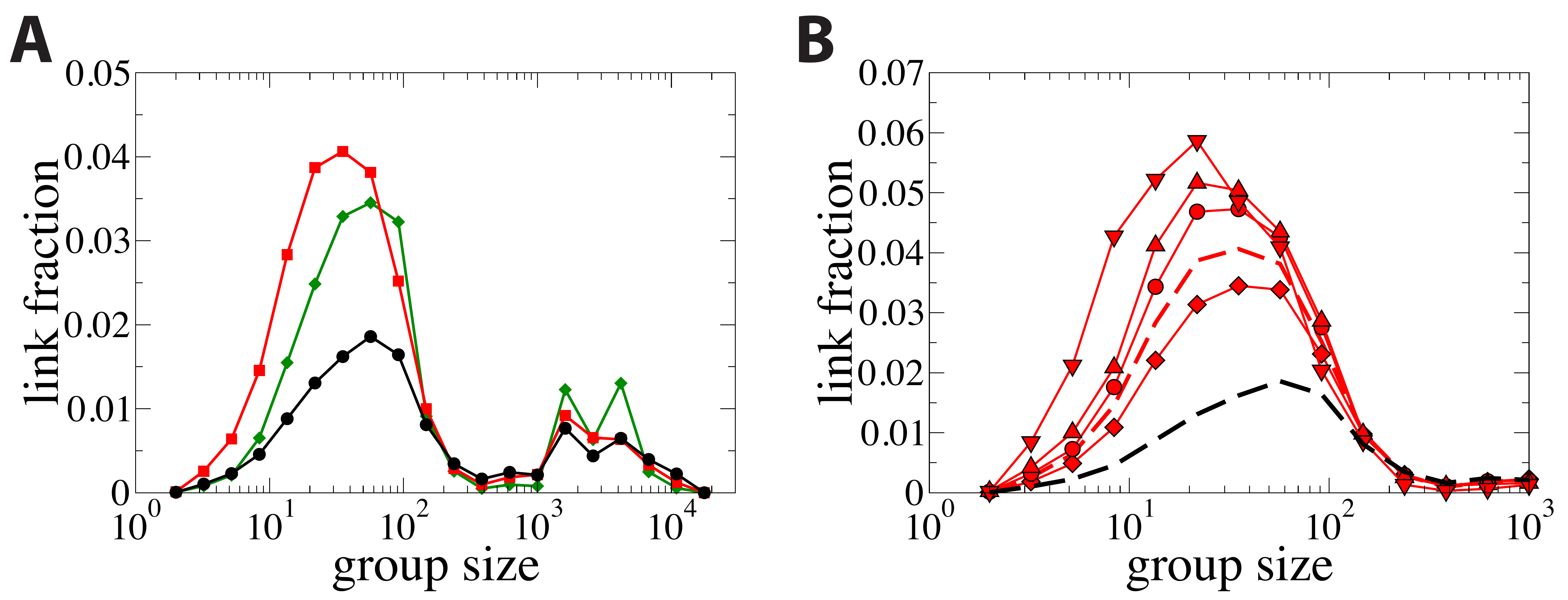}
\caption{(A) Fraction of links in the follower network, of links with mentions and links with retweets for bridges as a function of the size of the group. This figure is equivalent to the Figure 3A of the main text but for bridges instead of pure internal links. (B)  Fraction of links with mention activity of different intensity. The dashed curves are the total for the follower network (black) and for the links with mentions (red). While the other curves correspond (from bottom to top) to fractions of links with: 1 non-reciprocated mention (diamonds), 3 mentions (circles), 6 mentions (triangle up) and more than 6 mentions (triangle down).}
\label{bridges2}\end{center}
\end{figure}

\begin{acknowledgments}
The authors would like to thank Sergio G\'omez, Andrea Lancichinetti and Ian Leung for making network-clustering software available and for their advice in its use. P.A.G. and J.J.R. acknowledge support from the JAE program of the CSIC. Funding was provided by the Spanish Ministry of Science through projects MODASS (FIS2011-24785) and MOSAICO (FIS2006-01485). 

\end{acknowledgments}

\end{document}